
\input harvmac

\def\np#1#2#3{Nucl. Phys. {\bf B#1} (#2) #3}
\def\pl#1#2#3{Phys. Lett. {\bf #1B} (#2) #3}
\def\prl#1#2#3{Phys. Rev. Lett. {\bf #1} (#2) #3}
\def\physrev#1#2#3{Phys. Rev. {\bf D#1} (#2) #3}

\def\prep#1#2#3{Phys. Rep. {\bf #1} (#2) #3}

\def\cmp#1#2#3{Comm. Math. Phys. {\bf #1} (#2) #3}
\def\vev#1{\langle#1\rangle}
\def\Tr{{\rm Tr ~}}

\def\tilde{\widetilde}
\def\inbar{\,\vrule height1.5ex width.4pt depth0pt}
\def\IB{\relax{\rm I\kern-.18em B}}
\def\IC{\relax\hbox{$\inbar\kern-.3em{\rm C}$}}
\def\ID{\relax{\rm I\kern-.18em D}}
\def\IE{\relax{\rm I\kern-.18em E}}
\def\IF{\relax{\rm I\kern-.18em F}}
\def\IG{\relax\hbox{$\inbar\kern-.3em{\rm G}$}}
\def\IH{\relax{\rm I\kern-.18em H}}
\def\II{\relax{\rm I\kern-.18em I}}
\def\IK{\relax{\rm I\kern-.18em K}}
\def\IL{\relax{\rm I\kern-.18em L}}
\def\IM{\relax{\rm I\kern-.18em M}}
\def\IN{\relax{\rm I\kern-.18em N}}
\def\IO{\relax\hbox{$\inbar\kern-.3em{\rm O}$}}
\def\IP{\relax{\rm I\kern-.18em P}}
\def\IQ{\relax\hbox{$\inbar\kern-.3em{\rm Q}$}}
\def\IR{\relax{\rm I\kern-.18em R}}
\font\cmss=cmss10 \font\cmsss=cmss10 at 7pt
\def\IZ{\relax\ifmmode\mathchoice
{\hbox{\cmss Z\kern-.4em Z}}{\hbox{\cmss Z\kern-.4em Z}}
{\lower.9pt\hbox{\cmsss Z\kern-.4em Z}}
{\lower1.2pt\hbox{\cmsss Z\kern-.4em Z}}\else{\cmss Z\kern-.4em Z}\fi}
\def\CM{\cal M}
\def\IGa{\relax\hbox{${\rm I}\kern-.18em\Gamma$}}
\def\IPi{\relax\hbox{${\rm I}\kern-.18em\Pi$}}
\def\ITh{\relax\hbox{$\inbar\kern-.3em\Theta$}}
\def\IOm{\relax\hbox{$\inbar\kern-3.00pt\Omega$}}

\Title{hep-th/9510222, EFI-95-68, WIS/95/27, RU-95-75}
{\vbox{\centerline{Chiral Rings, Singularity Theory}
\centerline{and Electric-Magnetic Duality}}}
\medskip
\centerline{\it D. Kutasov}
\smallskip
\centerline{Enrico Fermi Institute and}
\centerline{Department of Physics}
\centerline{University of Chicago}
\centerline{Chicago, IL 60637, USA}

\vskip .2in
\centerline{\it A. Schwimmer}
\smallskip\centerline
{Department of Physics of Complex Systems}
\centerline{Weizmann Institute of Science}
\centerline{Rehovot, 76100, Israel}

\vskip .2in
\centerline{\it N. Seiberg}
\smallskip\centerline
{Department of Physics and Astronomy}
\centerline{Rutgers University }
\centerline{Piscataway, NJ08855, USA}

\vglue .3cm

\bigskip

\noindent
We study in detail the space of perturbations of a pair of dual $N=1$
supersymmetric theories based on an $SU(N_c)$ gauge theory with an
adjoint $X$ and fundamentals with a superpotential which is polynomial
in $X$.  The equivalence between them depends on non-trivial facts about
polynomial equations, i.e.\ singularity theory.  The classical chiral
rings of the two theories are different.  Quantum mechanically there are
new relations in the chiral rings which ensure their equivalence.
Duality interchanges ``trivial'' classical relations in one theory with
quantum relations in the other and vice versa.  We also speculate about
the behavior of the theory without the superpotential.

\Date{10/95}

\newsec{Introduction and Summary.}

The recent progress in the understanding of the dynamics of
supersymmetric theories (for recent reviews and an extensive list of
references see
\nref\powerd{N. Seiberg, The Power of Duality -- Exact Results in 4D
SUSY Field Theories. hep-th/9506077, RU-95-37, IASSNS-HEP-95/46, to
appear in the Proc. of PASCOS 95, the Proc. of the Oskar Klein lectures,
and in the Proc. of the Yukawa International Seminar '95}%
\nref\lectures{K. Intriligator and N. Seiberg,
Lectures on Supersymmetric Gauge Theories and Electric-Magnetic Duality.
hep-th/9509066, RU-95-48, IASSNS-HEP-95/70.  To appear in the Proc.\ of
Trieste '95 spring school, TASI '95, Trieste '95 summer school, and
Cargese '95 summer school.}%
\refs{\powerd, \lectures})
uncovered the crucial role played by electric-magnetic duality
\ref\mo{C. Montonen and D. Olive, \pl {72}{1977}{117}; P. Goddard,
J. Nuyts and D. Olive, \np{125}{1977}{1}.}
in understanding the strong coupling dynamics.
In scale invariant theories like the $N=4$
\ref\dualnf{H. Osborn, \pl{83}{1979}{321}; A. Sen, hep-th/9402032,
\pl{329}{1994}{217}; C. Vafa and E. Witten, hep-th/9408074,
\np{432}{1994}{3}.}
and finite $N=2$ theories
\ref\swii{N. Seiberg and E. Witten, hep-th/9408099,
\np{431}{1994}{484}.}
duality provides two different descriptions of the same physical system
that are equivalent at all distance scales.  In asymptotically free
theories the underlying degrees of freedom are visible at short distance
and therefore the existence of different descriptions that are
equivalent at all scales (exact duality) is impossible. Nevertheless,
duality may be generalized to such theories
\ref\nati{N. Seiberg, hep-th/9411149, \np{435}{1995}{129}.},
relating different quantum field theories with the same long distance
behavior. When this long distance behavior is described by a non-trivial
superconformal quantum field theory the dual theories are in the same
universality class -- they flow to the same fixed point of the
renormalization group.  When one of these theories is infra-red free it
gives a simple description of the long distance physics of its strongly
coupled dual.

No proof of this duality is known but there is a lot of evidence
supporting it.  There are three kinds of independent tests:

\item{1.}
The two dual theories have the same global symmetries and the 't Hooft
anomaly matching conditions for these symmetries are satisfied.

\item{2.} The two theories have the same moduli space of vacua.  These
are obtained by giving expectation values to the first components of
chiral superfields.

\item{3.} The equivalence is preserved under
deformations of the theories by the $F$-components of chiral operators.
In particular the moduli spaces and chiral rings agree as a function of
these deformations.

It is important to stress that in every one of these tests the classical
theories are different and only the quantum theories become equivalent.
There is also a crucial difference in the physical interpretation of the
deformations of the two theories along the moduli space and by the
chiral operators. Often, when one theory is Higgsed and becomes weaker,
its dual is confining and becomes stronger.  This is one of the reasons
for interpreting the relation between these theories as
electric-magnetic duality.

The last two tests above are closely related.  The rings of chiral
operators can be thought of as functions on the moduli space ${\CM}_0$.
Hence, one might think that test two above implies test three.  However,
such a relation is not always simple.  When there are points on the
moduli space with extra massless particles the situation is more
involved.  Then the moduli space ${\CM}_0$ is constrained also by the
equations of motion of these particles.  As one adds sources to the
theory proportional to $F$-components of chiral operators, the
expectation values of the chiral fields can move away {}from ${\CM}_0$.
The simplest way to describe the situation in such a case is to use an
enlarged field space $\CM$ which includes all the fields (including
those massive fields which become massless at special points) and to
write a superpotential on $\CM$.  Then, the equations of motion derived
{}from the superpotential lead to relations in the ring.  These
relations depend on the parameters in the theory -- the sources.

In previously studied examples the structure of the chiral ring was
relatively simple. In particular it was determined to a large extent by
the symmetries.  In general, the structure of the chiral ring can be
quite involved.  One may describe the ring in terms of generators
satisfying certain relations.  The relations in the classical chiral
ring are consequences of the composite nature of the gauge invariant
chiral operators .  Quantum mechanically these relations can be modified
\ref\natii{N. Seiberg, hep-th/9402044, \physrev{49}{1994}{6857}}.
We will see that it may also happen that new relations in the chiral
ring appear quantum mechanically. Then, the classical chiral ring is
truncated, in some cases rather dramatically.

The purpose of this paper is to study some qualitative and quantitative
features of the duality of \nati\ in a class of examples that exhibit
the phenomena mentioned above and a rich duality structure which helps
in analyzing them. In the rest of this section we will
describe the models we will study and state the main results.
Derivations and many additional details appear in subsequent sections.

\subsec{The models}

\centerline{\it The electric theory}

\nref\dk{D. Kutasov,  hep-th/9503086, \pl{351} {1995} 230.}%
\nref\ks{D. Kutasov and A. Schwimmer,  hep-th/9505004, \pl{354} {1995}
315.}%
\nref\asy{O. Aharony, J. Sonnenschein and S. Yankielowicz,
 hep-th/9504113, \np{449} {1995} 509.}%
\nref\intdual{K. Intriligator, hep-th/9505051,
\np{448} {1995} 187.}%
\nref\berkooz{M. Berkooz, RU-95-29, hep-th/9505067.}%
\nref\rlmsspso{R. Leigh and M. Strassler, hep-th/9505088, \pl{356}
{1995} 492.}%
\nref\ilst{K. Intriligator, R. Leigh and M. Strassler, hep-th/9506148,
RU-95-38.}%

Consider a $G=SU(N_c)$ gauge theory with an adjoint field $X$ and $N_f$
quarks $Q^i$ and $\tilde Q_{\tilde i}$ ($i,\tilde i=1,...,N_f$) in the
fundamental and anti-fundamental representations of the gauge group,
respectively.  These theories are
still not fully understood (see however a discussion below in section
{\it 7}).  When a superpotential
\eqn\welec{W=\sum_{i=0}^k{s_i\over k+1-i} \Tr X^{k+1-i} }
is turned on, the dynamics simplifies. At first sight the fact that the
high order polynomials appearing in \welec\ can have any effect on the
physics is surprising. Indeed, the presence of these non-renormalizable
interactions seems irrelevant for the long distance behavior of the
theory, which will be our main interest below. Nevertheless, these
operators have in general strong effects on the infrared dynamics. They
are examples of operators that in the general theory of the
renormalization group are known as {\it dangerously irrelevant}\foot{ We
thank S. Shenker who pointed the relevance of this notion in this
context.}. Some comments about the properties of such operators appear
in Appendix A.

To simplify the analysis of \welec\ with a traceless matrix $X$, we may
view $X$ as an arbitrary matrix and represent the constraint by a
Lagrange multiplier term $\lambda \Tr X$ in the superpotential.
Physically, this amounts to adding two massive chiral fields to our
problem, $ \lambda $ and $\Tr X$.  Clearly, this does not affect the
long distance behavior.  Then we can shift $X$ by a term proportional to
the identity matrix $X_s=X + {s_1 \over s_0k}1 $ to set the coefficient
of $\Tr X_s^k$ in \welec\ to zero; $\lambda$ is also shifted by a
suitable constant.  Such a shift removing the first subleading term in
the superpotential is a standard manipulation in singularity theory.
Rewriting the superpotential \welec\ in terms of the shifted $X$
corresponds to performing an analytic reparametrization on the space of
coupling constants. The new electric coupling constants will be denoted
by $\{t_i\}$. The explicit coordinate transformation {}from $\{s_i\}$ to
$\{t_i\}$ will appear below.

To find the classical moduli space of the theory one should first impose
the D-flatness equations and mod out by gauge transformations.  This is
equivalent to moding out the space of chiral fields by $SU(N_c)_{\IC}$.
Using this symmetry we can diagonalize $X$ and then impose the equation
of motion {}from \welec.  The eigenvalues of $X$ must satisfy
$W^\prime(x)=0$. For generic couplings $\{s_i\}$ there
are $k$ distinct solutions $c_1,\cdots, c_k$.  Vacua of the gauge theory
are labeled by sequences of integers $(r_1, r_2, \cdots, r_k)$; $r_l$ is
the number of eigenvalues of the matrix $\langle X\rangle$ residing in
the $l$'th minimum of the potential.  The gauge group is broken by the
$X$ expectation value:
\eqn\dd{SU(N_c)\to SU(r_1)\times SU(r_2)\times\cdots\times SU(r_k)
\times U(1)^{k-1}}
At low energies the theory describes $k$ decoupled supersymmetric
QCD (SQCD) systems\foot{The
different SQCD systems are in general coupled by high dimension operators
which are sometimes important.} with
gauge groups $SU(r_l)$ and gauged baryon number. For a given choice of
$\{r_l\}$ there is a moduli space of vacua associated with giving
expectation values to the quarks. Therefore, the classical moduli space
consists of many disconnected components parametrized by the $\{r_i\}$.
One can fine tune the couplings such that some of the eigenvalues
$\{c_i\}$ coincide. The resulting multicritical behavior will be
analyzed in section {\it 2}.

Quantum mechanically, not all vacua are stable. If, for example,
we pick a classical vacuum \dd\ with one or more of the $r_l>N_f$,
the resulting SQCD theory is destabilized by quantum effects
\nref\ads{I. Affleck, M. Dine, and N. Seiberg, \np{241}{1984}{493};
\np{256}{1985}{557}}%
\nref\cern{D. Amati, K. Konishi, Y. Meurice, G.C. Rossi and G.
Veneziano, \prep{162}{1988}{169} and references therein.}%
\refs{\ads,\cern}.  Hence, such classical vacua are not present in the
quantum moduli space.  Similarly, some of the vacua in the multicritical
case are destabilized and are removed {}from the quantum moduli space.

The classical chiral ring\foot{Setting the quark fields to zero for the
moment; the full structure will appear in section {\it 2}.} can be
thought of as the ring generated by the operators $\Tr X^j$ subject to
two classes of constraints. The first comes from
the equation of motion which follows {}from the superpotential \welec,
$W^\prime=0$.  The second comes {}from the characteristic polynomial of
$X$, and is an example of a relation following {}from the composite
nature of gauge invariant operators mentioned above.

Quantum mechanically one expects on general grounds
to find new relations in the
chiral ring corresponding to the quantum reduction of the moduli space
described above. It is in principle possible to construct these
relations by requiring that imposing them has the effect of removing
exactly the vacua that we know {}from our previous discussion should be
removed, but this is very difficult in practice as well as unmotivated.
Duality provides an elegant general solution to the problem, explaining
why such new relations appear in the quantum chiral ring and providing a
constructive way of determining them.

\bigskip
\centerline{\it The magnetic theory.}
\medskip
It was shown in \refs{\dk, \ks} that in the presence of a superpotential
\welec\ there exists a simple dual magnetic description\foot{For related
work see \refs{\asy - \ilst}.}. It is similar to the original electric
theory but based on the gauge group $SU(\bar N_c=kN_f-N_c)$ with an
adjoint field $Y$ and $N_f$ quarks $q_i$ and $\tilde q^{\tilde i}$ in
the fundamental and anti-fundamental representations as well as some
gauge invariant fields $(M_j)^i_{\tilde i}$ which correspond to the
composite operators
\eqn\defM{(M_j)^i_{\tilde i}=\tilde Q_{\tilde i} X_s^{j-1} Q^i;\;
j=1,\cdots, k }
where the suppressed color indices are summed over (when $s_1 \not= 0$
we find it convenient to define $M_j$ in terms of $X_s$).  The magnetic
theory has a superpotential
\eqn\wmagz{W_{\rm mag}=-\sum_l{ t_l\over k+1-l}\Tr
Y_s^{k+1-l}+{1\over\mu^2}\sum_{l=0}^{k-1} t_l\sum_{j=1}^{k-l}M_j\tilde q
Y_s^{k-j-l}q+\alpha_s(t)}
where the shifted field $Y_s$ is defined similarly to $X_s$ so that the
coefficient of $\Tr Y_s^k$ in the magnetic
superpotential vanishes.  The auxiliary scale $\mu$ is
needed for dimensional reasons; note that
even though the fields $M_j$ are elementary in the magnetic description,
the identification \defM\ implies that they are assigned scaling
dimension $j+1$.  Therefore, $\mu$ in \wmagz\ has indeed dimensions of
mass. One could redefine $M_j$ by powers of $\mu, s_0$ to make their
kinetic terms canonical. The numerical coefficients of the various terms
in \wmagz\ can be calculated using flows, and will be derived below. The
function $\alpha_s(t)$ will be computed too.

The magnetic superpotential \wmagz\ can be used to find the operator map
relating the operators $\Tr Y_s^j$ to the electric ones.
Differentiating the generating functional with respect to
the couplings $t_{k+1-j}$ one can derive the map of the operators
\eqn\mapops{{1 \over j} \Tr X_s^j = {\partial W \over \partial
t_{k+1-j}} = {\partial W_{\rm mag} \over \partial t_{k+1-j}} \quad.}
The function $\alpha_s$ in \wmagz\ is important because it contributes to
the map of the operators \mapops.

Note that \wmagz\ describes the duality map in the coordinates on theory
space described above, $\{t_i\}$. $Y_s$ is the adjoint field in those
coordinates (see section {\it 3}). The relatively simple form of the map
is in fact the main motivation behind the coordinate transformation
$s\to t$. In the original $s$ variables the duality map is significantly
more complicated. Its precise form will be exhibited below.

The discussion of the classical and quantum moduli space and chiral ring
for the theory \wmagz\ is essentially identical to the electric case,
replacing $N_c, r_l\to \bar N_c, \bar r_l$.  The classical electric and
magnetic moduli spaces are different; they have different numbers of
disconnected components (labeled by $\{r_i\}$, $\{\bar r_i\}$
respectively).  Similarly, the classical chiral rings are different.
Since the matrices $X$, $Y$ are of different size, the effects of the
characteristic polynomial are different in the two cases.

Quantum mechanically, the moduli spaces agree. After removing the
subspaces of moduli space corresponding to unstable vacua we find,
rather remarkably, the same number of components on both sides, with the
same physical properties. Some of these properties will be investigated
below.

Duality suggests a natural candidate for the quantum deformation of
the chiral rings in the two theories. It is very natural to add to the
classical relations coming {}from the equation of motion and the
characteristic polynomial in the electric theory an additional set of
relations following {}from the {\it magnetic} characteristic polynomial
using the duality map \mapops. These relations would be described as
quantum strong coupling effects in the electric theory (we will see that
they are trivial when the electric theory is weakly coupled), while in
the magnetic language they correspond to classical `trivial' relations
following {}from compositeness of $\Tr Y^j$. Similarly, the magnetic
chiral ring is modified by quantum relations obtained via the duality
map {}from the electric characteristic polynomial.  One of our goals in
this paper will be to establish that the quantum modification of the
chiral ring just described does indeed take place.

The detailed map between the electric theory and the magnetic theory
\wmagz, \mapops\ satisfies a number of non-trivial constraints:

\item{1.} The electric theory has a complicated vacuum structure which
depends on the details of the superpotential \welec.  The magnetic
theory should have the same vacuum structure.

\item{2.} The expectation values of the operators $\Tr X^j$ can be
calculated in all the vacua of the electric theory.  They should agree
with the corresponding calculation done using the magnetic variables and
the map of operators \mapops.

\item{3.} As various perturbations are turned on, some fields become
massive and can be integrated out both in the electric and in the
magnetic theories.  The two resulting low energy theories are
calculable.  They should be dual to one another. In particular, powerful
constraints arise {}from requiring consistency of the operator map
implied by duality with various deformations.

Additional constraints on duality follow {}from the requirement that
certain scale matching relations are consistent with all deformations.
These relations are discussed in the next subsection.

It is highly non-trivial and surprising that there exists a duality
transformation which satisfies such a large number of consistency checks.

\subsec{Scale matching.}

An important chiral operator in the electric theory is the kinetic term
of the gauge fields $W_\alpha^2$.  Its coefficient is $P \log \Lambda$
where $\Lambda$ is a dynamical scale related to the gauge coupling
by dimensional transmutation,
and $P$ is the coefficient of the one loop beta function (in an
$SU(N_c)$ gauge theory with $N_f$ flavors of
quarks in the fundamental representation
$P=3N_c-N_f$).  Even though    $W_{\alpha}$
is chiral (annihilated by $\bar D_{\dot
\alpha}$), it is usually not a chiral primary field at the IR fixed
point.
The reason is that the anomaly equation often\foot{This happens whenever
the superpotential $W$ vanishes, or more generally when $W$ is
independent of at least one of the matter superfields (which is the case
in the class of theories considered here).}
relates it to
another primary field $\CO$ as $W_\alpha^2=\bar D^2 \CO$.  Therefore it
is not in the chiral ring.  Nevertheless, its coefficient in the
electric Lagrangian should be related to its counterpart in the magnetic
theory, $\bar P \log \bar \Lambda$ ($\bar P$ is the coefficient
of the one loop beta function in the magnetic theory).  In our case
we will show that the relation has the form

\eqn\simrela{\Lambda ^{2N_c-N_f} \bar \Lambda ^{2\bar N_c-N_f}= \left(
{\mu\over s_0}\right)^{2N_f} .}
The relation \simrela\ shows in a quantitative way how one theory
becomes stronger as the other becomes weaker.  Its
consistency with various flows will provide rather stringent
quantitative tests of duality.

$\Lambda$ ($\bar\Lambda$) is usually thought of as the scale at which
the behavior of the electric (magnetic) theory crosses over {}from being
dominated by the short distance fixed point to the long distance one
(the typical mass scale of the theory). This raises a number of
questions; it is not clear why $\Lambda$, $\bar\Lambda$ defined in such
a way should satisfy a relation like \simrela\ when the theories are
only equivalent in the extreme infrared. Also, this definition leaves
ambiguous a numerical factor, especially in theories with more than one
scale.

One can think about eq. \simrela\ and the scales appearing in it purely
in the extreme IR theory. Since the scaling dimensions of various
operators in the infrared are not the same as in the UV (see \nati\ and
discussion below), one needs a dimensionful parameter to relate the UV
operators to the IR ones. That dimensionful parameter, which can be
defined for example through two point functions of such operators, is
$\Lambda$.  Similarly, in the magnetic theory one has $\bar\Lambda$, and
additional parameters such as $\mu$ \wmagz. The meaning of the
scale matching relation is that $\bar\Lambda$ and $\mu$
must be chosen to obey \simrela\ in order for the
correlation functions of the two theories to agree including
normalization.  At any given point in the space of theories we can
absorb the scales $\Lambda$, $\bar\Lambda$, $\mu$ into the definitions of
the operators, thus making the scale matching relation \simrela\ seem
trivial. The non-trivial content in \simrela\ is its consistency with
duality under all possible deformations of both theories. Indeed, we
will see that  consistency leads to highly non-trivial checks of
duality.  The situation is reminiscent of the Zamolodchikov metric in
two dimensional field theory, which is trivial at any given point in the
space of theories, but whose curvature carries invariant geometrical
information. The scale matching also describes an invariant relation
between the geometries of the spaces of electric and magnetic theories.

\subsec{Outline}

We plan to discuss two main issues. The first is the structure of the
quantum chiral ring and moduli space, and their transformation under
duality. The second is non-trivial quantitative tests of duality, which
are possible because of the large number of vacua that the system
possesses in general. In all these tests one uses symmetries to write
down certain duality relations.  This leaves in general some
undetermined functions of the coupling constants. These functions can be
calculated by assuming duality in some vacua of the theory. Since there
are in general many more vacua (and independent tests)
than unknown functions, the agreement of
the resulting structures with duality in all vacua is a non-trivial
check.

In section {\it 2} we review the results of \refs{\dk,\ks}. We describe
the class of theories we will study, describe the duality map, and
discuss their classical and quantum chiral rings and moduli spaces of
vacua. In particular we establish the existence of new quantum relations
in the chiral rings of these theories corresponding to classical
relations in the duals.

In section {\it 3} we construct the detailed map of the superpotential
\wmagz.  We describe the transformation of coordinates $\{s\}\to \{t\}$
and show that the duality map is rather simple in the $\{t\}$
coordinates. After constructing this map we change coordinates back to
the physically natural ones.

In section {\it 4} we turn to the gauge coupling constant matching
relation
\simrela.  We show that it is preserved under the various deformations.
In particular, with arbitrary coupling constants $t_i$ one can compute
the effective Lagrangian of the low energy theory both in the electric
and in the magnetic variables.  These lead (generically) to dual pairs
of supersymmetric QCD like theories whose scales are related by the
appropriate scale matching relation. The couplings $s_i$ can also be
fine tuned to yield multicritical infrared behavior, whose consistency
with duality puts further constraints on the structure described in
sections {\it 2, 3}.

In section {\it 5} we study the baryon operators in the theory and show
that they are mapped correctly between the electric and magnetic
theories.  This provides additional non-trivial checks of duality and
the explicit coefficients in the magnetic superpotential.

In section {\it 6} we illustrate the general results with a few
examples. In particular, we show that in some cases the quantum
relations in the chiral ring lead to qualitative changes in the
structure of the chiral ring.  We conclude with some comments about the
theory with no superpotential in section {\it 7}. Two appendices contain
a discussion of dangerously irrelevant operators and some useful
identities about polynomial equations which are used in the text.

\newsec{Supersymmetric Yang-Mills theory coupled to adjoint and
fundamental matter.}

\subsec{The electric theory.}

We start with a review of the results of \refs{\dk, \ks} on
supersymmetric Yang-Mills theory with gauge group $G=SU(N_c)$ coupled to
a single chiral matter superfield $X_\alpha^\beta$ in the adjoint
representation of the gauge group, and $N_f$ flavors of fundamental
representation superfields, $Q^i_\alpha$, $\tilde Q_j^\beta$; $\alpha,
\beta=1,\cdots, N_c$; $i,j=1,\cdots, N_f$. This theory is in a
non-Abelian Coulomb phase for all $N_f\ge1$.  Its anomaly free global
symmetry is
\eqn\glsym{SU(N_f)\times SU(N_f) \times U(1)_B \times
U(1)_{R_1}\times U(1)_{R_2} }
The two $SU(N_f)$ factors act by unitary transformations on $Q$, $\tilde
Q$ respectively; baryon number assigns charge $+1(-1)$ to $Q$ ($\tilde
Q$), while under the $R$ symmetries the superspace coordinates
$\theta_\alpha$ are assigned charge $1$, $Q$, $\tilde Q$ charge $B_f$,
and $X$ charge $B_a$; anomaly freedom implies that:
\eqn\an{N_fB_f+N_cB_a=N_f.}
Without a superpotential this model is not currently understood beyond
the vicinity of $N_f\simeq 2N_c$ where perturbative techniques are
reliable
\ref\bankszaks{T. Banks and A. Zaks, \np{196}{1982}{189}}.

One of the main points of \refs{\dk, \ks} was that the theory simplifies
if we add a superpotential\foot{In \ks\ $s_0/(k+1)$ was denoted by
$g_k$.}
\eqn\b{W={s_0\over k+1}\Tr X^{k+1}.}
This superpotential corresponds, for generic $k$, to a dangerously
irrelevant perturbation of the theory with $W=0$ (see Appendix A), and
thus cannot be ignored despite being irrelevant near the (free) UV
fixed point of the theory.  This superpotential has the effect of
truncating the chiral ring of the theory, imposing the constraint
\eqn\h{\left(X^k\right)_\alpha^\beta-{1\over N_c}
(\Tr X^k)\delta_\alpha^\beta={\rm D\;\; term}}
which follows {}from the equation of motion for $X$; it also removes
many of the flat directions of the original theory.  In addition, the
superpotential \b\ breaks one of the two $R$ symmetries in \glsym.  It
is useful to think of $s_0$ (and other couplings to be introduced below)
as background superfields whose lowest components get expectation
values
\ref\nonren{N. Seiberg, hep-ph/9309335, \pl{318}{1993}{469}}.
The superfield $s_0$ is then seen to transform under the
$U(1)_R$ symmetries \an\ with charge $B_0=2-(k+1)B_a$. Therefore, only
the $U(1)_R$ symmetry under which the charge of the adjoint field $X$ is
$B_a=2/(k+1)$ leaves the vacuum with $\langle s_0\rangle= s_0$
invariant, and corresponds to a good symmetry. The unbroken global
symmetry of the model is:
\eqn\globsym{SU(N_f)\times SU(N_f) \times U(1)_B \times U(1)_R }
The matter fields $Q$, $\tilde Q$ and $X$ transform under
this global symmetry as follows:
\eqn\d{\eqalign{
Q &\qquad (N_f,1,1, 1-{2\over k+1}{N_c\over N_f}) \cr
\tilde Q & \qquad (1, \overline N_f,-1,1-{2\over k+1}{N_c\over N_f})\cr
X &\qquad (1,1,0, {2\over k+1}) .\cr
}}

We will also be interested in more general superpotentials\foot{It is
standard in singularity theory to resolve a multicritical singularity
such as \b\ in order to study its properties.}:
\eqn\wpert{ W=\sum_{i=0}^{k-1}{s_i\over k+1-i}
{\rm Tr} X^{k+1-i}+\lambda {\rm Tr} X.}
$\lambda$ is a Lagrange multiplier enforcing the condition ${\rm Tr}
X=0$.  The coupling constant $s_i$ has dimension $2-k+i$ and $U(1)_R$
charge $B_i=2-(k+1-i)B_a$. For non zero $\{s_i\}$ \wpert\ breaks both $R$
symmetries \an.  The space of theories labeled by the $\{s_i\}$ describes
rather rich dynamics.  Using transformations in $SU(N_c)_{\IC}$ one can
diagonalize the matrix $X$.  For generic $\{s_i\}$, such that
\eqn\wprime{ W^\prime(x)=\sum_{i=0}^{k-1}s_i
x^{k-i}+\lambda \equiv s_0\prod_{i=1}^k(x-c_i)}
with all eigenvalues $c_i$ different {}from each other, the theory
splits in the infrared into a set of decoupled SQCD
theories.  Ground states are labeled by sequences of integers $r_1\le
r_2\le\cdots\le r_k$, where $r_l$ is the number of eigenvalues of the
matrix $X$ residing in the $l$'th minimum of the potential
$V=|W^\prime(x)|^2$. Clearly, $\sum_{l=1}^kr_l=N_c$.  $\lambda$ in
\wpert\ is determined by requiring that the sum of the eigenvalues
(which depends on $\lambda$) vanishes,
\eqn\trcond{\sum_{i=1}^k c_i r_i=0.}

In each vacuum $X$ has a quadratic superpotential, i.e.\ it is massive
and can be integrated out. The gauge group is broken by the $X$
expectation value:
\eqn\dd{SU(N_c)\to SU(r_1)\times SU(r_2)\times\cdots\times SU(r_k)
\times U(1)^{k-1}}
$SU(r_l)$ is the gauge group of the $l$'th decoupled SQCD theory one
finds in the infrared\foot{Some of the $r_l$ may vanish, in which case
\dd\ is modified in an obvious way.}. The existence of a large number of
vacua \dd\ (labeled by partitions of $N_c$ into $k$ or less integers)
after resolving the singularity \wpert\ is the key fact leading to
simple but non-trivial quantitative checks of duality.

For generic $\{s_i\}$ (and fixed $\{r_i\}$) the classical infrared
behavior of the theory, a decoupled set of SQCD theories with gauge
groups \dd, is insensitive to the precise values of $\{s_i\}$.
Quantum mechanically , this is not the case;
each of the low energy SQCD theories has a scale
$\Lambda_i$, $i=1,\cdots, k$. These $k$ scales are
functions of the scale $\Lambda$ of the underlying theory with the
adjoint superfield, and the $k-1$ couplings $s_1, \cdots, s_{k-1}$. The
functions $\Lambda_i(\Lambda, s_i)$ will be computed below (in section
{\it 4}).

A similar discussion holds if the $\{s_l\}$ are fine
tuned such that some of the $c_i$ in \wprime\ coincide:
\eqn\wwprime{ W^\prime(x)=\sum_{i=0}^{k-1}s_i {\rm Tr}
X^{k-i}+\lambda = s_0\prod_{i=1}^m(x-c_i)^{n_i}}
where $\sum_i n_i=k;\;\;m\le k;\;\; n_i\ge1$.  Vacua are labeled by
sequences $(r_1,\cdots, r_m)$, $\sum_i r_i=N_c$, corresponding to
different ways of partitioning the $N_c$ eigenvalues among the $m$
critical points $c_i$ \wwprime. In this case one finds in the infrared a
decoupled set of theories with gauge groups $SU(r_l)$ and
superpotentials of the form \b\ with $k=n_l$ (see
\wwprime). Therefore, the deformations \wpert\ connect theories with
different $k$ in \b\ and unify them into a single framework.

It is often possible to deduce non-trivial
properties of the theory \b\ by studying the
deformed theories \wpert. As a simple example
\ks, turning on small $s_i$ in \wpert\ and
using the fact that the resulting SQCD theories
have stable vacua iff all $r_i\le N_f$ one deduces
that the theory \b\ has a vacuum iff
\eqn\ee{N_f\ge{N_c\over k}.}
We will see other examples below.

\subsec{The magnetic theory}

The main result of \refs{\dk, \ks} was that the strongly coupled
infrared physics of theory \b\ can be studied using a dual description
in terms of a ``magnetic'' supersymmetric gauge theory with gauge group
$\bar G=SU(\bar N_c)$, $\bar N_c=kN_f-N_c$, and the following matter
content: $N_f$ flavors of (dual) quarks, $q_{\bar i}^{\bar\alpha}$,
$\tilde q^{\bar j}_{\bar \beta}$, an adjoint superfield
$Y_{\bar\alpha}^{\bar\beta}$, and gauge singlets $(M_j)_{\tilde i}^i$
representing the generalized mesons,
\eqn\mj{(M_j)_{\tilde i}^i=
\tilde Q_{\tilde i} X^{j-1}Q^i;\;\;\;j=1,2,\cdots, k}
of the original, ``electric'' theory. The mesons $M_j$
have in the magnetic
theory standard kinetic  terms $\int d^4\theta
M_j^\dagger M_j$, rescaled by powers
of $s_0$, $\mu$.
The magnetic superpotential is:
\eqn\f{W_{\rm mag}={\bar s_0\over k+1}\Tr Y^{k+1}+
{s_0\over\mu^2}\sum_{j=1}^k M_j\tilde q Y^{k-j} q.}
The auxiliary scale $\mu$ (mentioned
in the introduction) is needed for dimensional reasons .
In the next sections we will see that it is very natural to normalize
$Y$ such that $\bar s_0=-s_0$;  this choice leads
to all the coefficients\foot{Which as we will see are uniquely
determined by consistency of duality with deformations.}
in the sum over $j$ in \f\ being $1$, as indicated.

The transformation properties of the magnetic matter fields,
$q$, $\tilde q$, $Y$ and $M_j$ under the global symmetries \globsym\
are:
\eqn\e{\eqalign{
q &\qquad (\overline N_f,1,{N_c\over kN_f-N_c}, 1-{2\over
k+1}{kN_f-N_c\over
 N_f})
\cr
\tilde q & \qquad (1, N_f,-{N_c\over kN_f-N_c},1-{2\over k+1}{
kN_f-N_c\over N_f}) \cr
Y &\qquad (1,1,0, {2\over k+1}) \cr
M_j &\qquad (N_f,\overline N_f,0, 2-{4\over k+1}{N_c\over
N_f}+{2\over k+1}(j-1)) \cr
}}

The auxiliary scale $\mu$ in \f\ is actually not an independent
parameter.  The scale of the electric theory $\Lambda$, that of the
magnetic theory $\bar \Lambda$, and $\mu$ satisfy the scale matching
relation described in the introduction:
\eqn\scmat{\Lambda^{2N_c-N_f}\bar\Lambda^{2\bar N_c-N_f}=
Cs_0^{-2N_f}\mu^{2N_f}}
It is easy to check that \scmat\ is invariant under all global
symmetries including those under which $s_0$, $\mu$, $\Lambda$
transform. It is in fact uniquely fixed by these symmetries.  As in
\nref\isson{K. Intriligator and N. Seiberg,
hep-th/9503179, \np{444}{1995}{125}.}
\refs{\isson,\lectures}, the scale matching condition \scmat\ implies
that when the electric theory is weakly coupled the magnetic one is
strongly coupled, and vice versa\foot{As an example, the electric theory
is weakly coupled when $N_f$ is slightly below $2N_c$, and free for
$N_f\ge 2N_c$.  The magnetic theory is strongly coupled in that whole
region.}. Differentiating the actions with respect to
$\Lambda$ holding $s_0$ and $\mu$ fixed, the electric gauge field
strength $W_\alpha^2$ is related to the magnetic one $\bar W_\alpha^2$
by the relation:
\eqn\wa{\bar W_\alpha\bar W^\alpha=-W_\alpha W^\alpha.}
It is highly non-trivial, and will be shown in section {\it 4}, that the
scale matching relation \scmat\ which is completely determined by global
symmetries, is consistent with all possible deformations of the theory,
such as turning on masses for the quarks $Q^i$, giving expectation
values to the mesons $M_j$ \mj, and deforming the theory to non-zero
$s_i$ \wpert.  The numerical constant $C$ in \scmat\ is also fixed by
the flows, $C=1$, and its value and in particular its independence of
$N_f$, $N_c$, $k$ lead to additional tests of duality.

We view the consistency of \scmat\ with deformations as strong evidence
for the validity of the electric-magnetic duality hypothesis of
\refs{\nati, \dk, \ks}.

When the electric superpotential is deformed to \wpert\ the electric
theory develops a large number of vacua labeled by ($r_1$, $\cdots$,
$r_m$), with $\sum r_i=N_c$ (see the discussion after \wwprime). The
$SU(r_l)$ theory contains an adjoint field with superpotential $W\simeq
X^{n_l+1}$.  The magnetic theory has a similar structure obtained by
analyzing vacua of the superpotential \wmagz. Its vacua are labeled by
integers ($\bar r_1, \cdots, \bar r_m$), $\sum \bar r_i=\bar N_c$.
Classically, the moduli spaces do not agree. However \refs{\dk, \ks}
when we include the quantum stability constraints \ee\ we find a one to
one correspondence of the quantum vacua in the two theories. The vacuum
map is non trivial:
\eqn\vacmap{(r_1,r_2,\cdots, r_m)\leftrightarrow(n_1N_f-r_1, n_2
N_f-r_2,\cdots, n_mN_f-r_m).}

A class of operators that was discussed in \refs{\dk, \ks}, and will be
revisited in section {\it 5}, are the baryon-like operators:
\eqn\baryon{\eqalign{&B^{i_1\cdots i_{n_1},j_1\cdots
j_{n_2},\cdots, z_1\cdots z_{n_k}}=\cr
&\epsilon^{\alpha_1\cdots\alpha_{n_1},\beta_1
\cdots\beta_{n_2},\cdots, \rho_1\cdots\rho_{n_k}}
Q^{i_1}_{\alpha_1}\cdots Q^{i_{n_1}}_{\alpha_{n_1}}
(XQ)^{j_1}_{\beta_1}\cdots (XQ)^{j_{n_2}}_{\beta_{n_2}}\cdots
(X^{k-1}Q)^{z_1}_{\rho_1}\cdots (X^{k-1}Q)^{z_{n_k}}_{\rho_{n_k}}\cr}}
where $\sum_i n_i=N_c$. The duality map relating the baryons
$B$ \baryon, and the analogously
defined dual baryons $\bar B$ is:
\eqn\bmap{\eqalign{&B^{i_1\cdots i_{n_1},j_1\cdots j_{n_2},\cdots,
z_1\cdots z_{n_k}}=P
\left(\prod_{i=1}^k{1\over \bar n_i!}\right)
{\bar s_0}^{kN_f\over2}\left(\mu^2\right)^{-{\bar N_c\over 2}}
\Lambda^{{k\over2}(2N_c-N_f)}\cr
&\epsilon^{i_1\cdots i_{n_1},\bar z_1\cdots
\bar z_{\bar n_k}}\epsilon^{j_1\cdots j_{n_2},
\bar y_1\cdots\bar y_{\bar n_{k-1}}}
... \epsilon^{z_1\cdots z_{n_k},\bar i_1\cdots
\bar i_{\bar n_1}}
\bar B_{\bar i_1\cdots \bar i_{\bar n_1},\cdots,
\bar y_1\cdots \bar y_{\bar n_{k-1}},
\bar z_1\cdots \bar z_{\bar n_k}}\cr}
}
where $\bar n_l=N_f-n_{k+1-l}$, $l=1,\cdots, k$, and $P$ is a phase that
will be discussed in section {\it 5}.  The form of \bmap\ is determined
by global symmetries, while the overall numerical constant is uniquely
fixed by the flows, and it is non-trivial that it is consistent with the
deformations, such as \wpert. This compatibility will be established in
section {\it 5}.

In addition to the mesons \mj\ and baryons \baryon, the chiral ring of
the electric (magnetic) theory contains the generators ${\rm Tr} X^j$
(${\rm Tr} Y^j$), $j=2,\cdots, k$.  Global symmetries require (up to
terms depending on the quarks which we will determine below) that
\eqn\trx{{\rm Tr} Y^j=f_j {\rm Tr} X^j}
with $f_j$ calculable numerical constants.  In the next section we will
calculate $f_j$
and describe the generalization of \trx\ to the deformed theories
\wpert.

The duality map described in this section may be used to study the
physics of the theory \b\ at strong coupling using a weak coupling
description in terms of magnetic variables. For example, one finds that
this theory may exhibit at strong coupling a ``free magnetic phase,''
with the magnetic variables governed by a non-asymptotically free gauge
theory.  Some additional features are described in \refs{\dk, \ks}.

\subsec{Classical and quantum chiral rings.}

The classical chiral ring of the electric theory is generated by the
generalized mesons $M_j$ \mj, baryons \baryon\ and traces of the adjoint
matrix, $\Tr X^j$.  The latter, which we will focus on here satisfy
classically two sets of constraints, following {}from the equation of
motion $W^\prime=0$, \wprime\ and {}from the characteristic polynomial.
Defining $f(p)\equiv\det(p-X)$, the constraint is $f(X)=0$. The two
together give rise to a set of algebraic equations for the generators of
the classical chiral ring.  The solutions of these equations describe
the ring of functions on the classical moduli space.

One can repeat the same construction for the magnetic theory. The
discussion proceeds in complete parallel to that of the moduli space.
Classically the chiral rings of the electric and magnetic theories
cannot be the same: while the equation of motion $W^\prime=0$ gives
similar relations for the generators, the characteristic polynomials give
different sets of constraints, since the sizes of the matrices $X$ and
$Y$ are different. However, the discussion of the correspondence of the
quantum moduli spaces points to the correct modification of the chiral
rings in the quantum theory needed to restore duality and the relation
between the (quantum) chiral rings and moduli spaces.  For generic
values of the couplings $\{s_i\}$ the quantum restriction on the
$\{r_i\}$, $r_l\le n_l N_f$ turns under duality \vacmap\ to the trivial
condition $\bar r_l\ge 0$ (and vice versa). This means that if we add
the {\it classical} characteristic polynomial
relations on the magnetic chiral ring as {\it
quantum} relations in the electric theory, we are guaranteed that they
will have precisely the right effect on the electric moduli space,
eliminating the unstable electric vacua and leaving all other parts of
the electric moduli space unchanged.

To summarize, we have the following structure of the electric and
magnetic chiral rings. Both are generated by the operators $\Tr X^j$ (or
$\Tr Y^j$ -- the two sets of operators are related by the transformation
\mapops\ which we will make more explicit below) satisfying three sets
of constraints:

\item {1.} The equation of motion, $W^\prime=0$.

\item {2.} The vanishing of the electric characteristic polynomial,
$f_{\rm el}=0$.

\item {3.} The vanishing of the magnetic characteristic polynomial,
$f_{\rm mag}=0$.

The first set of relations appears classically in both the electric and
magnetic theories and gives similar constraints in both.  The second
set of relations is classical in the electric theory (and is due to the
compositeness of $\Tr X^j$), but it is a non-trivial quantum effect in
the magnetic theory. The third is classical in the magnetic theory and
quantum in the electric one.

\bigskip
\newsec{The mapping of the superpotential.}

To complete the discussion of the previous section we must construct
the duality map taking the electric theory \b\ to the magnetic one
\f\ when the electric theory is deformed to \wpert. One expects
the magnetic superpotential \f\ to be deformed as well, and in this
section we will discuss the detailed way in which this happens.  We will
start with a discussion of the deformation of the first term on the
r.h.s. of \f, the superpotential for $Y$, and then turn to the second
term, proportional to $M_j$. In the process we will verify all the
numerical coefficients in \wmagz\ and learn some qualitative things
about the duality map. We start with a discussion of the $Y$
superpotential.

\subsec{The problem.}

The electric superpotential \wpert\ describes a space of theories
parametrized by the couplings $s_i$.
On general grounds one expects a magnetic superpotential
\eqn\bwpert{ \bar W=\sum_{i=0}^{k-1}{\bar s_i\over k+1-i}
{\rm Tr} Y^{k+1-i}+ \bar\lambda {\rm Tr} Y +\alpha(s).}
$\bar s_i=\bar s_i(s)$ are the magnetic coupling constants and
$\alpha(s)$ is a constant.  The purpose of this section is to find $\bar
s(s)$ and $\alpha(s)$.

In complete analogy with the discussion of the electric theory in
section {\it 2}, the magnetic theory \bwpert\ exhibits for generic $\bar
s_i$ a large number of vacua, parametrized by integers $\bar r_l$
corresponding to the number of eigenvalues of the matrix $Y$ with the
value $\bar c_l$, the $l$'th minimum of the magnetic superpotential
$|\bar W^\prime|^2$. The $\{ \bar c_l\}$ are defined by a magnetic
analogue of \wprime.  Clearly $\sum_l\bar r_l= \bar N_c=kN_f-N_c$. The
low energy magnetic theory is a direct product of decoupled copies of
SQCD with $N_f$ flavors of quarks, with the gauge group broken according
to:
\eqn\ddd{SU(\bar N_c)\to SU(\bar r_1)\times
SU(\bar r_2)\times\cdots\times SU(\bar r_k) \times U(1)^{k-1}}

The picture proposed in \ks\ was that the original duality between \b\
and \f\ reduces for the deformed theories \wpert, \bwpert\ to a direct
product of the SQCD dualities of \nati\ for the separate factors in \dd,
\ddd.  That means that the magnetic multiplicities $(\bar r_1,\cdots,
\bar r_k)$ are related to the electric ones via the SQCD duality
relation (compare to \vacmap):
\eqn\vacrel{\bar r_i=N_f-r_i.}
The fact that \vacrel\ is a one to one map of the sets of vacua of
the electric and magnetic theories follows {}from results of
\refs{\ads, \cern} on vacuum stability in SQCD.

Furthermore, it was argued in \ks\ that when two or more of the critical
points of $W$ coincide, as in \wwprime, the same number of critical
points of $\bar W$ should coincide.  If, using the notation of equation
\wwprime, the order of a critical point $c_i$ (and therefore that of
$\bar c_i$ as well) is $n_i$, the degeneracies $r_i$ and $\bar r_i$ of
this critical point in the electric and magnetic theories are related by
\eqn\vva{\bar r_i=n_iN_f-r_i.}
The duality of section {\it 2} induces in this case a duality of a
similar kind, between an electric theory with gauge group $SU(r_i)$ and
a superpotential $\Tr X^{n_i+1}$, and a magnetic one with gauge group
$SU(\bar r_i)=SU(n_iN_f-r_i)$ with a similar superpotential.

For the above scenario to be realized, the electric and magnetic
superpotentials must be closely related. In particular, the fact that
whenever any number of critical points $c_i$ coincide, the same number
of dual critical points $\bar c_i$ must coincide as well is a very
strong constraint on the dual couplings $\bar s_i(s)$. A naive guess for
the solution would be a proportionality relation between the electric
and magnetic couplings,
\eqn\wrong{\bar s_i=cs_i}
with $c$ a constant.  Throughout this paper we will be using the
convention
\eqn\conv{\bar s_0=-s_0}
which defines the normalization of $Y$ relative to $X$. This convention
would set $c=-1$. Eq. \wrong\ implies that $\bar c_i=c_i$ and
automatically satisfies the degeneration of singularities requirement
described above.  However, the mapping $\bar s_i(s)$ cannot (in
general) be as simple as \wrong\ because of the non-trivial mapping of
electric to magnetic multiplicities, \vacrel. Indeed, in a vacuum with
given $\{r_l\}$, the tracelessness of $X$ implies that $\sum_{l=1}^k
c_lr_l=0$, whereas, assuming $\bar c_i=c_i$ and using \vacrel, the
tracelessness of $Y$ is the condition $\sum_{l=1}^k c_l(N_f-r_l)=0$.
The two are incompatible unless $s_1$ in \wpert\ vanishes. For non zero
$s_1$ we conclude that the mapping $\bar s(s)$ must be non trivial; we
will construct it below.

The duality map relating $\Tr X^j$ to $\Tr Y^l$ is closely related to
the mapping $\bar s(s)$.  Define the free energy of the model as
\eqn\frr{e^{-\int d^4x d^2 \theta  F(s_i) + c.c}=\langle e^{- \int d^4x
d^2\theta W(X, s_i)+ c.c.}\rangle}
where $s_i$ are background chiral superfields.  Then, correlation
functions of the operators ${\rm Tr} X^j$ are given by derivatives of
the free energy $F$ with respect to the superfields $s_i$:
\eqn\expec{{1\over k+1-i}\langle{\rm Tr} X^{k+1-i}\rangle=
{\partial F\over\partial s_i}}
and similarly in the magnetic theory, in terms of the dual
free energy $\bar F$:
\eqn\bexpec{{1\over k+1-i}\langle{\rm Tr} Y^{k+1-i}\rangle=
{\partial \bar F\over\partial \bar s_i}}
where duality implies
\eqn\tuv{\bar F(\bar s_i(s))=F(s_i).}
Since $\Tr X^j$, $\Tr Y^j$ are tangent vectors to the space of theories,
\expec, \bexpec\ we find that the electric and magnetic operators are
related by:
\eqn\integ{{1\over k+1-i}{\rm Tr} X^{k+1-i}=\sum_j
{\partial\bar s_j\over\partial s_i}
{1\over k+1-j}{\rm Tr} Y^{k+1-j}+{\partial\alpha\over\partial s_i}}
Taking the expectation values of both sides of eq. \integ\ we find that
the mapping $\bar s_i(s)$ must satisfy a constraint in addition to the
previously described one on the degeneration of eigenvalues. The
expectation values of the left and right hand sides of \integ\ which
depend in a highly non-trivial way on the particular vacuum chosen (the
set of $r_i$) must satisfy a relation that is {\it independent} of the
particular vacuum chosen.  Clearly, the special form of the
mapping \integ\ and the large number of vacua in which it should hold
presents a formidable constraint on their form.

\subsec{A reparametrization of the space of theories and a general
solution.}

It is convenient to think
of $X$, $Y$ as general $U(N)$ matrices, with a dynamical Lagrange
multiplier $\lambda$ ($\bar\lambda$) imposing the tracelessness of $X$
($Y$). Consider the electric theory, described by \wpert.  It is
convenient to define a shifted $X$, denoted by $X_s$ as:
\eqn\guru{X_s\equiv X+b{\bf 1}}
with
\eqn\shiff{b={s_1\over s_0 k}.}
The shift \guru, \shiff\ cancels the first subleading term in $W$,
leading to the superpotential:
\eqn\gpert{W_s(X_s)=\sum_{i=0}^{k-1}{t_i\over
k+1-i}\Tr X_s^{k+1-i}+\lambda_s\left(\Tr X_s-bN_c\right)+\beta N_c}
where $W_s(X_s)\equiv W(X)$,
\eqn\slag{\eqalign{t_i=&\sum_{j=0}^i{k-j\choose i-j}(-b)^{i-j}s_j\cr
                    \lambda_s=&\lambda+\sum_{j=0}^{k-1}(-b)^{k-j}s_j\cr
                    \beta=&-\sum_{j=0}^{k-1}{k-j\over
                    k+1-j}(-b)^{k+1-j}s_j.\cr}}
Note that $t_0=s_0$ and $t_1=0$.  The transformation \guru, \shiff\
corresponds to an analytic coordinate transformation on the space of
theories \wpert.  A similar transformation can be performed on \bwpert,
replacing $Y$ by $Y_s$ and $\bar\lambda$ with $\bar\lambda_s$.  The
$k-1$ independent couplings $s_i$, $i=1,\cdots, k-1$ are replaced by the
$k-2$ couplings $t_i$, $i=2,\cdots, k-1$, and $b$ \shiff. In the $X_s$
variables the coefficient of the first subleading term $\Tr X_s^k$ in
the superpotential \gpert\ always vanishes. The information about that
coefficient in the original description \wpert\ is in $b$ \shiff.  In a
sense, the transformation \guru, \shiff\ allowed us to trade the
operator $\Tr X^k$ for the operator $\lambda$, which is possible since
by the $X$ equation of motion (ignoring $D$ terms as usual)
\eqn\ring{\lambda={1\over N_c}\sum_{i=0}^{k-2}s_i\Tr X^{k-i}.}

Since in the $t_i$, $b$ parametrization of the space of theories the
first subleading term in $W$ vanishes by construction, it is natural to
postulate that the duality map for the eigenvalues of $X_s$ , $a_i$
defined analogously to \wprime\ is simply
\eqn\ansatz{\bar a_i= a_i;\;i=1,\cdots k.}
With the convention \conv\ this means that (compare to \wrong):
\eqn\rel{\eqalign{\bar t_i=&-t_i\cr
                   \bar\lambda_s=&-\lambda_s.\cr}}
The second equation in \rel\ is an operator identity; it can be thought
of as arising {}from the coupling relations:
\eqn\rels{\bar b\bar N_c=- bN_c;\;\;\alpha_s={\rm independent\;of\;} b}
using \integ.  Equations \rel, \rels\ specify the mapping $\bar s_i(s)$
completely.

We now determine $\alpha_s$.  Using \integ, \rel\ we find the operator
equation
\eqn\cond{\Tr {X_s}^{k+1-i}=-\Tr
{Y_s}^{k+1-i}+(k+1-i){\partial\alpha_s\over\partial t_i}}
($i=2,\cdots, k-1$).  The expectation values of the l.h.s of \cond\ in a
vacuum specified by a set of $\{r_l\}$ is
\eqn\xtrace{\Tr {X_s}^{k+1-i}=\sum_{l=1}^kr_l{a_l}^{k+1-i}}
while in the magnetic theory:
\eqn\xtrace{\Tr {Y_s}^{k+1-i}=\sum_{l=1}^k\bar r_l{\bar
a_l}^{k+1-i}
=\sum_{l=1}^k(N_f-r_l){a_l}^{k+1-i}=
-\Tr {X_s}^{k+1-i}+N_f u_{k+1-i}}
where (see also Appendix B):
\eqn\defu{u_j\equiv\sum_{i=1}^k{a_i}^j.}
Comparing \cond\ and \xtrace\ we see that for consistency of the picture
advocated above we must be able to write the $u_j$ as:
\eqn\totder{u_{k+1-j}={k+1-j\over
N_f}{\partial\alpha_s\over\partial t_j}}
Indeed, one can check that eq. \totder\ is satisfied
with $\alpha_s$ given by:
\eqn\formm{\alpha_s={N_f\over k+1}\sum_{i=2}^{k-1}
it_i{u_{k+1-i}\over k+1-i}}
The proof of the fact that \formm\ satisfies \totder\ uses the following
property of the $u_n$ \defu:
\eqn\knbig{{\partial\over\partial t_j}{u_{j+l}\over j+l}={\rm
independent\;of}\;j.}
This and other properties of the $u_n$ are reviewed in Appendix B.

Summarizing, the main results of this subsection are the mapping of the
couplings in the electric superpotential $t_i$, $b$ defined by \slag,
\shiff\ to their magnetic counterparts, given by equations \rel, \rels.
This simple transformation law induces a transformation \cond\ for the
operators $\Tr X_s^j$ with $j=2, \cdots, k-1$. The operator $\Tr X^k$
that was conjugate to the coupling $s_1$ is eliminated in favor of the
Lagrange multiplier $\lambda$ which also has a simple transformation
given by the second equation in \rel.

Of course, the simple transformation laws described in this subsection
become  more complicated when we translate them back to the
original coordinates $s_i$.

\subsec{The duality map in the original variables.}

After the discussion of the previous subsection it is not difficult to
describe the duality map for the perturbed superpotential \wpert\ in the
original coordinates $s_i$. The main point is that while as we shall see
the mapping $\bar s_i(s)$ (and therefore the operator map \integ) is
somewhat complicated in this case, the mapping of the eigenvalues
remains simple. Indeed, using the simple relation between the
eigenvalues of $X_s$ and of $Y_s$ \ansatz\ and the relation between $X$
and $Y$ and their shifted counterparts $X_s$ and $Y_s$ \guru, \shiff\ we
conclude that the mapping of the eigenvalues of $X$, $Y$, ($c_i$, $\bar
c_i$) is:
\eqn\oansatz{\bar c_i=c_i+d;\;\; d\equiv{s_1 N_f\over s_0\bar N_c}.}
Similarly, we derive the map of the coupling constants
\eqn\co{\bar s_m=-\sum_{l=0}^ms_{m-l}(-d)^l{k-m+l\choose l}}
and the operators (using \integ):
\eqn\k{\Tr Y^j=-\sum_{i=2}^j{j\choose i}
d^{j-i}\Tr X^i+N_f\bar u_j-N_c d^j;\;\;\; j<k}
\eqn\kk{\eqalign{\Tr Y^k=&-\sum_{i=2}^k{k\choose i} d^{k-i}\Tr
X^i+{kN_f\over N_c} \sum_{j=0}^{k-2} {s_j\over s_0} \Tr X^{k-j}\cr
+&\bar N_cd^k+ N_f\sum_{j=1}^{k-1}{k\choose
j}d^{k-j}u_j-N_f\sum_{l=1}^{k-1} {s_l\over s_0} u_{k-l}.\cr}}

The function $\alpha$ defined in \integ\ can be expressed in terms of
$\beta$ and $\alpha_s$  defined in \slag, \formm\ as:
\eqn\nnn{\alpha(s_i)=\beta N_c-\bar\beta\bar N_c+\alpha_s(t_i(s)).}

Interestingly, when all $s_i$ except $s_0$ vanish (i.e.\
the superpotential is \b) there is nevertheless a non-trivial operator
matching following {}from equations \k, \kk:
\eqn\xspecial{\eqalign{\Tr Y^j=&-\Tr X^j;\;j=2,\cdots, k-1\cr
                      \Tr Y^k=&{\bar N_c\over N_c}\Tr X^k\cr}}
{}From eq. \xspecial\ one can read off the values of the coefficients
$c_j$ of section {\it 2} \trx.

\subsec{The $M_j$ terms in the magnetic superpotential.}

So far our discussion focused on the way the first term in the magnetic
superpotential \f\ is deformed as we deform the electric superpotential
\wpert. In this subsection we will use these results to determine the
deformation of the second term in $W_{\rm mag}$. We will work in the
parametrization of the space of theories described in subsection {\it
3.2}.

When one turns on non-vanishing couplings $t_i$ in \wpert, the magnetic
superpotential \f\ can in principle receive contributions
proportional to $t_j$, $t_jt_l$, etc, consistently with the global
symmetries. The way to fix all these terms is to require that duality
act in the way described after eq. \ddd. Namely, for generic $t_i$ we
expect the magnetic theory to split into an approximately {\it
decoupled} set of SQCD theories that are dual to the different decoupled
factors in \dd.

This requirement of decoupling is rather non-trivial since the second
term on the r.h.s. of \f\ tends to couple the different $SU(\bar r_i)$
theories. Indeed, denote the first $r_1$ components (in color) of
the electric quarks\foot{In this subsection flavor indices will be
suppressed.}\ $Q$ by $Q_1$, the next $r_2$ by $Q_2$ and so on.
Similarly, the first $\bar r_1$ components of $q$ are denoted by $q_1$,
the next $\bar r_2$ by $q_2$, etc..  Then expanding around $\vev{X_s}$
we find
\eqn\mjj{M_j=\tilde Q_1 Q_1 a_1^{j-1}+\tilde Q_2 Q_2 a_2^{j-1}+\cdots
+\tilde Q_k Q_k a_k^{j-1}}
(recall that for generic $s_i$ we defined $M_j=\tilde Q X_s^{j-1}Q$)
$\tilde Q_l Q_l$ are the mesons of the $l$'th electric SQCD
theory with gauge group $SU(r_l)$. The color $SU(r_l)$ indices are
as usual suppressed and summed over. Similarly we write:
\eqn\qjj{\tilde q Y_s^j q=\tilde q_1 q_1 a_1^j+\tilde q_2 q_2
a_2^j+\cdots +\tilde q_k q_k a_k^j}
where as in {\it 3.2} we denote the shifted $Y$ field appropriate for
the $\bar t_i$ coordinate system on theory space by $Y_s$.
In the above formula we used the fact that in the coordinates
$t_i$ the duality map is trivial, $\bar a_i=a_i$ \ansatz.

The second term in $W_{\rm mag}$ \f\ has to be corrected in such a way
that the different SQCD theories do not couple -- there should not be
any cross terms coupling $\tilde q_jq_j$, $\tilde Q_i Q_i$ with $i\not
=j$.  The unique solution to this requirement is
\eqn\newwm{W_{\rm mag}=\sum_l{\bar t_l\over k+1-l}\Tr
Y_s^{k+1-l}+{1\over\mu^2}\sum_{l=0}^{k-1} t_l\sum_{j=1}^{k-l}M_j\tilde q
Y_s^{k-j-l}q.}
All the numerical coefficients in the second term on the r.h.s. of
\newwm\ are fixed by the requirement that when we substitute \mjj, \qjj\
into it, cross terms such as $\tilde Q_1 Q_1\tilde q_2 q_2$ vanish.
Indeed, the coefficient of the above operator is proportional to:
\eqn\eqzer{\sum_{l=0}^{k-1}t_l
\sum_{j=1}^{k-l}a_1^{j-1}a_2^{k-j-l}=\sum_{l=0}^{k-1}
t_l{a_2^{k-l}-a_1^{k-l}\over a_2-a_1}=0}
which vanishes because $a_1$, $a_2$ are roots of $W^\prime$ (see
\wprime).
Thus, with the choice of couplings in \newwm\ the magnetic theory
reduces
for generic $t_i$ into {\it decoupled} SQCD theories as required by
duality\foot{When $Y$ is integrated out there can be more terms of
higher dimension in the low energy superpotential which we do not
discuss.}.

The form \newwm\ which at this stage of the discussion is completely
fixed, must satisfy additional consistency conditions. The simplest of
these involves getting the right behavior when some of the roots $a_i$
coincide. For example, if $a_1=a_2$ and all other $a_i$ are different,
\mjj\ is replaced by:
\eqn\mjjnew{M_j=a_1^{j-1}\tilde Q_1 Q_1 +
(j-1)a_1^{j-2}\tilde Q_1X_{s1}Q_1+a_3^{j-1}\tilde Q_3 Q_3 +\cdots
+a_k^{j-1}\tilde Q_k Q_k .}
with $X_{s1}$ the fluctuating deviation of the shifted adjoint field of
$SU(r_1)$ {}from its v.e.v. Similarly, \qjj\ is replaced by:
\eqn\qjjnew{\tilde q Y_s^j q=a_1^j\tilde q_1 q_1 +
ja_1^{j-1}\tilde q_1Y_{s1}q_1+a_3^j\tilde q_3 q_3 +\cdots+
a_k^j\tilde q_k q_k .}
Using \mjjnew, \qjjnew\ in \newwm\ we find the correct superpotential
for decoupled SQCD and an $SU(\bar r_1)$ sector with $k=2$
as required by the duality.
All these checks give the expected results.  More generally,
when some eigenvalues $a_i$ coincide as in \wwprime, one finds that the
terms that must vanish are always proportional to derivatives of $W$ at
$a_i$ which vanish.

Additional consistency conditions on the detailed form of \newwm\ will
appear in the next section. We see again that consistency of the deformed
theory with duality fixes uniquely coefficients in the superpotential
of the unperturbed theory \f.

It is also useful to note at this point that the $t_l$ dependence of the
magnetic superpotential \newwm\ implies a correction to the dual of $\Tr
X_s^j$ given in \cond.  Indeed, differentiating the free energies of the
electric and magnetic theories (see \mapops, \frr\ -- \integ) one finds:
\eqn\condd{\Tr {X_s}^{k+1-i}=-\Tr {Y_s}^{k+1-i}+
{k+1-i\over\mu^2}\sum_{j=1}^{k-i}M_j\tilde q Y^{k-j-i}_s q+
(k+1-i){\partial\alpha_s\over\partial t_i}}
($i=2,\cdots, k-1$).  The second term on the r.h.s. mixes the operators
$\Tr X_s^j$ with the generalized magnetic mesons.

\newsec{Consistency of scale matching with deformations.}

In section {\it 2} we mentioned the relation \scmat\ between the scale
of the electric theory, $\Lambda$, that of the magnetic theory, $\bar
\Lambda$, and the dimensionful parameter $\mu$ entering \f, \newwm:
\eqn\bscmat{\Lambda^{2N_c-N_f}\bar
\Lambda^{2\bar N_c-N_f}=s_0^{-2N_f}\mu^{2N_f}}
It is interesting to check whether this relation is consistent with the
various deformations that the model possesses. These include adding
terms proportional to $M_j$ \mj\ to the superpotential, and turning on
$s_i$ \wpert. In this section we will check the compatibility of
\bscmat\ with two kinds of flows:

\item{1.} Adding a mass term to one of the flavors (e.g.
$m(M_1)_{N_f}^{N_f}$).

\item{2.} The general $s_i$ perturbations.

\noindent
In doing that we should stress that the relation \bscmat\ does not
depend on the masses or $s_i$.  This follows from the symmetries.

We will see that, remarkably, there is complete detailed agreement of
the two kinds of flows with \bscmat.   We start with a summary of the
conventions we will be using.

\subsec{Conventions.}

Before discussing the flows we must specify the threshold corrections
relating the scales of the theory when massive particles are integrated
out. When we integrate out a massive fundamental chiral superfield $Q$
with mass $m$, the $SU(N_c)$ gauge theory with an adjoint and $N_f$
flavors goes in the infrared to one with an adjoint and $N_f-1$ flavors,
and:
\eqn\cone{\Lambda_{N_c,N_f}^{2N_c-N_f}m=\Lambda_{N_c,
N_f-1}^{2N_c-(N_f-1)}}
The analogous relation in SQCD is:
\eqn\ctwo{\Lambda_{N_c,N_f}^{3N_c-N_f}m=\Lambda_{N_c,
N_f-1}^{3N_c-(N_f-1)}}
When we integrate out a chiral adjoint field $X$ of mass $m$, we have:
\eqn\cthree{m^{N_c}\Lambda_{N_c, N_f}^{2N_c-N_f}=\Lambda_{N_c,N_f}^{3N_c
-N_f}}
where the scale on the l.h.s. corresponds to the theory with an adjoint,
and the scale on the r.h.s to SQCD.  Finally, when we integrate out a
massive vector superfield of mass $m$ in the fundamental representation
of the gauge group (i.e.\ when part of the gauge group is Higgsed) we
have in the theory with an adjoint and $N_f$ fundamentals:
\eqn\cfour{\Lambda_{N_c, N_f}^{2N_c-N_f}=m^2
\Lambda_{N_c-1, N_f}^{2N_c-N_f-2}}
whereas in SQCD:
\eqn\cfive{\Lambda_{N_c, N_f}^{3N_c-N_f}=m^2
\Lambda_{N_c-1, N_f-1}^{3N_c-N_f-2}}
Notice that in the theory with the adjoint \cfour\ $N_f$ does not
decrease under Higgsing.  This is because one massless chiral superfield
is eaten by the massive gauge field but another appears {}from
decomposing the adjoint of $SU(N_c)$ w.r.t. $SU(N_c-1)$.

Out of \cone\ -- \cfive\ only three definitions are independent (e.g.
\cone, \cthree, \cfour). With these conventions the scale matching
condition in SQCD is \lectures:
\eqn\sqcd{\Lambda_{\rm SQCD}^{3N_c-N_f}\bar
\Lambda_{\rm SQCD}^{3\bar N_c-N_f}= (-)^{N_f-N_c}\mu^{N_f}}
Here $\mu$ is an auxiliary scale similar to that in \f.  It is defined
such that the magnetic superpotential in SQCD is :
\eqn\wmagsqcd{W_{\rm mag}^{\rm SQCD}={1\over\mu} M\tilde q q.}
With the conventions in hand we next turn to examine the flows.

\subsec{The mass flow.}

Consider adding to the electric theory a mass term:
\eqn\Qmass{W_{\rm el}={s_0\over k+1}\Tr X^{k+1}+m\tilde Q_{N_f}Q^{N_f}.}
The theory loses a flavor in the infrared; the scales of the high and
low energy theories ($\Lambda_{N_c, N_f}$ and $\Lambda_{N_c, N_f-1}$
respectively) are related by \cone:
\eqn\elsc{\Lambda_{N_c, N_f}^{2N_c-N_f}={1\over m}
\Lambda_{N_c, N_f-1}^{2N_c-(N_f-1)}.}
In the magnetic theory, the superpotential \f\ is modified to:
\eqn\fnew{W_{\rm mag}={\bar s_0\over k+1}\Tr Y^{k+1}+
{s_0\over\mu^2}\sum_{j=1}^k M_j\tilde q Y^{k-j} q+m(M_1)^{N_f}_{N_f}.}
One next needs to set the massive fields to solutions of their equations
of motion and integrate them out.  It is easy to see that all the fields
$(M_j)_i^{N_f}$, $(M_j)_{N_f}^i$ ($i=1,\cdots, N_f$), and components of
$q$, $\tilde q$, $Y$ in $k$ of the $\bar N_c$ directions are massive.
The expectation values of $\tilde q^{N_f}$, $q_{N_f}$, $Y$ satisfy:
\eqn\expval{\tilde q^{N_f}Y^{l-1}q_{N_f}=-
\delta_{l,k}{m\mu^2\over s_0};\;\;l=1,\cdots, k}
Taking into account the $D$ terms (which fix the relative normalization
of $\tilde q, q, Y$) and $Y$ equation of motion leads to:
\eqn\expvalb{\eqalign{\tilde q^{N_f}_\alpha=&
\delta_{\alpha,1}\left({m\mu^2\over\bar s_0}
\right)^{1\over k+1};\cr
q_{N_f}^\alpha=&\delta^{\alpha,k}\left({m\mu^2\over\bar s_0}
\right)^{1\over k+1};\cr
Y^\alpha_\beta=&\cases{\delta^\alpha_{\beta-1}\left({m\mu^2\over\bar s_0}
\right)^{1\over k+1}&$\beta=1,\cdots, k$\cr
                        0& otherwise\cr}\cr}}
We can think of the effect of $m$ in \fnew\ in two stages.  First, the
magnetic gauge group $SU(\bar N_c)$ is broken by the Higgs mechanism
to $SU(\bar N_c-k)$. At this stage, $q_{N_f}$, $\tilde q^{N_f}$,
$Y^\alpha_m$ $m=2,\cdots, k$, $Y_\alpha^s$ $s=1,\cdots, k-1$
($\alpha=k+1,\cdots\bar N_c$) gain a mass and join $k$ massive vector
superfields in the fundamental representation of the unbroken, $SU(\bar
N_c-k)$ gauge group. According to our convention \cfour\ this generates
a factor of the $k$'th power of the mass squared of the vector
superfields, $\left[\left({m\mu^2\over\bar s_0}\right)^{2\over
k+1}\right]^k$.

In a second stage, $Y_1^\alpha$ and $Y_\alpha^k$ get a mass {}from
expanding the superpotential:
\eqn\massY{W_{\rm mag}={\bar s_0\over k+1}\Tr Y^{k+1}\simeq
\bar s_0\langle Y\rangle^{k-1}
Y_1^\alpha Y^k_\alpha=\bar s_0\left({m\mu^2\over\bar s_0}
\right)^{k-1\over k+1}Y_1^\alpha Y^k_\alpha.}
In \massY\ we have used the fact that in expanding $\Tr Y^{k+1}$ to
leading order in $Y_1^\alpha Y^k_\alpha$ we must take these two $Y$'s
next to each other; terms like $\Tr \langle Y^n\rangle Y_1 \langle
Y^m\rangle Y^k$ ($n,m\not=0$) do not contribute such mass terms.  Since
$Y_1^\alpha$, $Y_\alpha^k$ can be thought of as massive chiral
superfields in the fundamental representation of $SU(\bar N_c-k)$ with
mass $\bar s_0\left({m\mu^2\over\bar s_0}
\right)^{k-1\over k+1}$ (see \massY, \expvalb),
we use \cone\ for the scale matching.

Finally, combining the two stages we have:
\eqn\scbar{\bar\Lambda_{\bar N_c, N_f}^{2\bar N_c-N_f}=
\left({m\mu^2\over\bar s_0}
\right)^{2k\over k+1}{1\over \bar s_0\left({m\mu^2\over\bar s_0}
\right)^{k-1\over k+1}}\bar\Lambda_{\bar N_c-k, N_f-1}^{2(\bar
N_c-k)-N_f} ={m\mu^2\over \bar s_0^2}\bar\Lambda_{\bar N_c-k,
N_f-1}^{2(\bar N_c-k)-N_f}}
Using  \elsc, \scbar\ we conclude that \bscmat\ leads to:
\eqn\nml{{1\over m}\Lambda_{N_c, N_f-1}^{2N_c-(N_f-1)}\;
{m\mu^2\over\bar s_0^2}\bar\Lambda_{\bar N_c-k, N_f-1}^{2(\bar
N_c-k)-N_f} =\mu^{2N_f} s_0^{-2N_f}}
or:
\eqn\nfmo{\Lambda_{N_c, N_f-1}^{2N_c-(N_f-1)}\bar
\Lambda_{\bar N_c-k, N_f-1}^{2(\bar
N_c-k)-N_f}=\mu^{2(N_f-1)}s_0^{-2(N_f-1)}}
which is exactly the right scale matching relation for the theory
with $N_f-1$ flavors. We therefore conclude that the scale matching
relation \bscmat\ is consistent with the mass perturbation \Qmass.

An interesting element of the preceding analysis is the fact that the
overall relative coefficient between the first and second terms in the
magnetic superpotential \f\ was important for quantitative
agreement of the matching conditions.  This relative coefficient was not
fixed by the discussion in section {\it 3.4} and we can view the
analysis of this subsection as a way to determine it.  We will soon see
a non-trivial check of its value {}from the $s_i $ flows \wpert.

\subsec{Deformation of the $X$ superpotential: the generic case.}

In this subsection we will be deforming the superpotential for $X$ ($Y$)
in the electric (magnetic) theory, as in \wpert, \bwpert.
Recall the duality
\eqn\simp{\eqalign{
W(X_s)=&\sum_{i=0}^{k-1}{t_i\over k+1-i}\Tr X_s^{k+1-i}\cr
\bar W(Y_s)=&\sum_{i=0}^{k-1}{\bar t_i\over k+1-i}\Tr Y_s^{k+1-i}
+\alpha_s \cr
\bar t_i=&-t_i;\;\bar a_i=a_i\cr}
}
where $a_i$ are the eigenvalues of $X_s$ defined as in \wprime\ and $\bar
a_i$ are similarly related to $Y_s$.

To test \bscmat\ in the deformed theory \simp\ we proceed in two stages.
First consider generic $t_i$ such that all eigenvalues $a_i$
\wprime\ are distinct.  Then the IR electric theory is a direct product
of decoupled SQCD theories with gauge groups $SU(r_i)$ (see discussion
following \wprime).  The scale of the $i$'th SQCD theory is related to
that of the high energy theory by:
\eqn\breaksqcd{\Lambda^{2N_c-N_f}=\Lambda_i^{3r_i-N_f}\left[
\prod_{j\not=i}(a_i-a_j)^{ 2r_j}\right]
{1\over \left[W^{\prime\prime}(a_i)\right]^{r_i}}}
One can think of \breaksqcd\ as following {}from a two step process.
First we turn on an expectation value for $X_s$:
\eqn\exX{\langle X_s\rangle={\rm diag}\left(a_1^{r_1},a_2^{r_2},
\cdots,a_k^{r_k}\right)}
where $a_1^{r_1}$ means the eigenvalue $a_1$ appears $r_1$ times, etc.
This makes some vector fields massive and using \cfour\ we get the
factor of $\left[\prod_{j\not=i}(a_i-a_j)^{ 2r_j}\right]$ in \breaksqcd.
Then, at a second stage, we integrate out the massive adjoint field in
the $SU(r_i)$ theory using \cthree.  This gives rise to ${1\over
\left[W^{\prime\prime}(a_i)\right]^{r_i}}$.

We evaluate $W^{\prime\prime}(a_i)$ using \wprime\ and find for
$\Lambda_i$:
\eqn\lami{\Lambda_i^{3r_i-N_f}=\Lambda^{2N_c-N_f}t_0^{r_i}\prod_{j\not=
i} (a_i-a_j)^{r_i-2r_j}}
 Repeating the same arguments for the magnetic theory
using \simp\ we find:
\eqn\blami{\bar\Lambda_i^{3\bar r_i-N_f}=\bar \Lambda^{2\bar N_c-N_f}
\bar t_0^{\bar r_i}\prod_{j\not= i}
(a_i-a_j)^{\bar r_i-2\bar r_j}.}

At this stage we have no independent check on \lami, \blami\ separately
(although one will appear in the next section when we discuss baryons),
but multiplying the left and right hand sides of \lami\ and \blami\ and
using \bscmat, \sqcd\ we find
\eqn\interm{\Lambda_i^{3r_i-N_f}\bar\Lambda_i^{3\bar r_i-N_f}=
(-)^{N_f-r_i}\mu_i^{N_f} =(-)^{N_f-r_i}{\mu^{2N_f}\over
t_0^{N_f}}\prod_{j\not=i}(a_i-a_j)^{-N_f}}
which means that
\eqn\mui{\mu_i={\mu^2\over t_0}{1\over\prod_{j\not=i}(a_i-a_j)}.}
Recall that $\mu_i$ is defined through the magnetic superpotential
in the $SU(\bar r_i)$ theory (compare to \wmagsqcd):
\eqn\wi{W_{\rm mag}^{(i)}={1\over\mu_i}\tilde Q_i Q_i\tilde q_i q_i.}
The scale parameters $\mu_i$ can be independently
calculated by the analysis
described in section {\it 3.4}.  By decomposing $M_j$, $\tilde q Y^l q$
into their $SU(r_i)$, $SU(\bar r_i)$ components \mjj, \qjj\ and
evaluating the coefficient of $\tilde Q_i Q_i \tilde q_i q_i$ in the
magnetic superpotential \newwm\ we find:
\eqn\calcmu{{1\over\mu_i}={1\over\mu^2}\sum_{l=0}^{k-1}t_l\sum_{j=1}^{k-l}
a_1^{j-1} a_1^{k-j-l}= \sum_{l=0}^{k-1}(k-l)t_la_1^{k-l-1}
={1\over\mu^2}W^{\prime\prime}(a_i)={t_0\over\mu^2}\prod_{j\not=i}
(a_i-a_j)}
which is exactly the right value \mui. Since the magnetic superpotential
\newwm\ is completely fixed by the considerations of section {\it 3},
the agreement between \mui\ and \calcmu\ gives another non-trivial
quantitative check of duality.

We conclude that, at least for generic $t_i$, the deformation \simp\ is
consistent with the scale matching relation \bscmat.  This still leaves
the question of whether we get a consistent picture when some $\{a_i\}$
coincide, which we briefly address in the next subsection.

\subsec{Deformation of the $X$ superpotential: coinciding eigenvalues.}

Consider for simplicity the special case discussed in section {\it 3.4},
where two of the $a_i$, say $a_1$ and $a_2$, coincide\foot{The
discussion can be trivially extended to the general case.}.  Then
instead of $k$ copies of SQCD, one gets in the IR $k-2$ copies of SQCD,
corresponding to $a_3,\cdots, a_k$, and one copy of the model
\b\ with $k=2$. The discussion of the $k-2$ SQCD vacua is exactly as in
the last subsection. The only new feature here is the discussion of the
scale of the $SU(r_1)$ $k=2$ model corresponding to $a_1$, and its dual,
the $SU(\bar r_1)$ model with $k=2$ and $\bar r_1=2N_f-r_1$.  The scale
$\Lambda_1$ of that low energy model is related to $\Lambda$ by:
\eqn\brktwo{\Lambda_1^{2r_1-N_f}=
\Lambda^{2N_c-N_f}\prod_{j=3}^k(a_j-a_1)^{-2r_j}}
by an argument analogous to that following \breaksqcd.  Similarly:
\eqn\bbrktwo{\bar\Lambda_1^{2\bar r_1-N_f}=\bar
\Lambda^{2\bar N_c-N_f}\prod_{j=3}^k(a_j-a_1)^{-2\bar r_j}}

Defining $t_0^{(l)}$ and $\mu^{(l)}$ to be the analogues of $t_0$, $\mu$
in the low energy theory, we have the scale matching relation:
\eqn\matchl{\Lambda_1^{2r_1-N_f}\bar\Lambda_1^{2\bar r_1-N_f}=
\left( t_0^{(l)}\right)
^{-2N_f}\left(\mu^{(l)}\right)^{2N_f}}
Multiplying \brktwo\ and \bbrktwo\ and using \bscmat, \matchl\ we find:
\eqn\somu{{t_0^{(l)}\over\mu^{(l)}}={t_0\over\mu}
\prod_{j=3}^k(a_j-a_1)}
One can again perform an independent check of \somu\
by calculating $t_0^{(l)}$, $\mu^{(l)}$
directly. Starting with $t_0^{(l)}$ we use \wprime\ near
$x\simeq a_1$:
\eqn\soo{W^\prime(x)=(x-a_1)^2\prod_{j=3}^k(x-a_j)\simeq t_0
\left[\prod_{j=3}^k
(a_1-a_j)\right] (x-a_1)^2\equiv t_0^{(l)}(x-a_1)^2}
so that:
\eqn\tol{t_0^{(l)}=t_0\prod_{j=3}^k(a_1-a_j).}
To get $\mu^{(l)}$ we use the decompositions \mjjnew, \qjjnew\ and
calculate the coefficient of $\tilde Q_1 Q_1\tilde q_1 Y_{s1} q_1$ (or
equivalently $\tilde Q_1 X_{s1}Q_1\tilde q_1 q_1$).  We find that
\eqn\mul{{t_0\over\mu^2}\prod_{i=1}^k(a_1-a_i)= {t_0^{(l)}
\over(\mu^{(l)})^2};\;
{\rm or:}\; \mu^{(l)}=\mu.}
Combining \tol, \mul\ we see that \somu\ is indeed valid and
the structure of flows \wpert\ is
consistent with the scale matching relation \bscmat.

\newsec{The mapping of baryons in the deformed theory.}

In section {\it 2} we proposed an exact mapping \bmap\ between the
baryon operators in the electric and magnetic theories. The mapping
involves powers of the dimensionful couplings that are fixed by global
symmetries and numerical factors that can be fixed by consistency with
the various flows. In this section we will outline the checks of
consistency of \bmap\ with various deformations. We will see that these
consistency checks lead to additional highly non-trivial checks of
duality. In particular, since $\Lambda$ appears in the map \bmap, we will
have a more direct check on the expressions for the scales $\Lambda_i$
\lami, \blami\ derived in section {\it 4}.

There is an inherent phase ambiguity in \bmap\ related to the
possibility of making field redefinitions in the two theories
corresponding to global symmetries of the problem.  Performing a baryon
number transformation in the magnetic theory introduces an
arbitrary phase in \bmap\ and an opposite phase in the relation between
$\tilde B$ and its dual $\bar {\tilde B}$. We can use this freedom to
make the phases in the mappings of $B$ and $\tilde B$ identical. This
still leaves a sign ambiguity in all our formulae for mapping of $B$,
$\tilde B$ separately, which disappears in the mapping of $B\tilde B$.
We will not write explicitly these $\pm$ signs, but instead leave the
branches of square roots appearing in $B$ and $\tilde B$ unspecified.
Phase factors appearing in the matching of $B\tilde B$ have an absolute
meaning and therefore will be calculated.

In the first subsection we will discuss the consistency of \bmap\ when
turning on masses for quarks \Qmass. In the second subsection we will
verify the consistency of \bmap\ with deformations of the superpotential
$W(X)$ \wpert.

\subsec{The mass flow.}

As in section {\it 4} we add a mass term for $Q^{N_f}$, $\tilde Q_{N_f}$
\Qmass.  The electric theory now has $N_f-1$ flavors. The electric
baryon \baryon\ splits into components with the indices
$i_{p_1},j_{p_2},\cdots, z_{p_k}$ taking values between $1$ and $N_f-1$
which stay massless, and the rest that become massive.  In the magnetic
theory we deform the superpotential by an $mM_1$ term \fnew. The gauge
group breaks {}from $SU(\bar N_c)$ to $SU(\bar N_c-k)$ and the number of
flavors decreases by one, $N_f\to N_f-1$. One notes that:

\item {1.} It follows {}from \expvalb\ that:
\eqn\expectqj{\langle\left(Y^jq^{N_f}\right)^\alpha\rangle=
\delta^\alpha_{k-j}\left(
{m\mu^2\over \bar s_0}\right)^{j+1\over k+1}}

\item {2.} All components of the magnetic quarks $q$, $\tilde q$
and the adjoint field $Y$ with color indices between $1$ and $k$, as
well as $q^{N_f}$, $\tilde q_{N_f}$ (all color components) become
massive.

\item {3.} Due to the structure of the r.h.s. of \bmap\ if any one or
more of the indices $i_{p_1}, \cdots, z_{p_k}$ equals $N_f$ the
corresponding magnetic baryon (the r.h.s. of \bmap) is massive, since it
is not possible to saturate the color indices $\bar\alpha=1,
\cdots, k$ corresponding to broken generators except with massive
quarks. This is in agreement with the behavior of the electric theory
(the l.h.s. of \bmap) as mentioned above.

\item {4.} If all indices $i_{p_1}, \cdots, z_{p_k}$ are less than $N_f$
the r.h.s. of \bmap\ does lead to massless baryons in the magnetic
theory.  The broken color indices $\bar\alpha=1, \cdots, k$ can now be
saturated by the expectation values of the $Y^jq^{N_f}$.  The only way
to saturate the broken color indices by the expectation values
\expectqj\ is to replace in $\bar B$ \bmap\ one of each group of indices
$\bar i, \cdots, \bar z$ by the appropriate expectation value in
\expectqj.  This leads to the following transformation of the r.h.s. of
\bmap\ under the mass flow:
\eqn\barflow{ \eqalign{&\epsilon^{i_1\cdots i_{n_1},\bar z_1\cdots
\bar z_{\bar n_k}}\epsilon^{j_1\cdots j_{n_2},
\bar y_1\cdots\bar y_{\bar n_{k-1}}}
... \epsilon^{z_1\cdots z_{n_k},\bar i_1\cdots \bar i_{\bar n_1}}
\bar B^{\bar N_c, N_f}
_{\bar i_1\cdots \bar i_{\bar n_1},\cdots, \bar y_1\cdots \bar y_{\bar
n_{k-1}},
\bar z_1\cdots \bar z_{\bar n_k}}\to (-)^{k(k-1)\over2}       \cr
&\bar n_1\bar n_2\cdots \bar n_k\left({m\mu^2\over \bar
s_0}\right)^{k\over 2}
\epsilon^{i_1\cdots i_{n_1},\bar z_1\cdots
\bar z_{\bar n_k-1}}\cdots
\epsilon^{z_1\cdots z_{n_k},\bar i_1\cdots \bar i_{\bar n_1-1}}
\bar B^{\bar N_c-k, N_f-1}
_{\bar i_1\cdots \bar i_{\bar n_1-1},\cdots,
\bar z_1\cdots \bar z_{\bar n_k-1}}\cr}}
The factors of $\bar n_i$ come {}from the number of possibilities of
placing the expectation value among the $\bar n_i$ $X^{i-1}Q$ operators.
The phase is due to a certain reordering of the $k$ color indices
$1,\cdots, k$ that is needed to bring the baryon into standard form
after symmetry breaking.  A similar factor does not appear in the
analogous mapping for $\tilde B$.  Clearly, the only meaningful quantity
is the phase appearing in the transformation of $B\tilde B$.

Inserting \barflow\ into \bmap\ we see that all the factors generated in
the breaking magically arrange to give the form \bmap\ again, with
$N_f\to N_f-1$. In particular, one notes that $\bar n_i\to \bar n_i-1$,
and the scale absorbs the factor of $m$ (see \cone).  Eq. \barflow\ and
the analogous relation following {}from   the analysis of the mapping
\bmap\ under turning on an expectation value for $M_k$ in the electric
theory allow one to determine the dependence on $N_f$ and $N_c$ of the
unambiguously defined phase in the mapping relation for $B\tilde B$.
One finds that the phase, called $P$ in \bmap\ is:
\eqn\phase{P^2=(-)^{{k(k-1)\over 2}N_f-N_c}.}
As discussed above, only $P^2$, which is a sign, is
meaningful.

\subsec{The baryon mapping in the presence of a deformed
superpotential.}

When a general superpotential \wpert\ is present the gauge group breaks
generically to \dd. Correspondingly, the baryon operator \baryon\ can be
expressed in terms of the baryons of the SQCD theories \dd. Similarly,
the baryon of the magnetic $SU(\bar N_c)$ theory is decomposable into
the baryons of the magnetic SQCD theories \ddd. The mapping of baryons
in the high energy theory, \bmap, should reduce in the deformed theory
to the baryon mapping in the individual SQCD theories:
\eqn\barguru{B^{i_1\cdots i_{N_c}}=
{1\over \bar N_c!}\Lambda^{{1\over2}(3N_c-N_f)}\bar
\mu^{{1\over2}(N_f-N_c)}
\epsilon^{i_1\cdots i_{N_c}
\bar i_1\cdots \bar i_{N_f-N_c}}\bar B_{\bar i_1\cdots \bar i_{\bar N_c}}
}
Eq. \barguru\ has the same sign ambiguity discussed above.

The general discussion of the reduction of \bmap\ under the deformations
\wpert\ is unfortunately rather complicated. To get a flavor of the
issues involved we will discuss here the special case of a theory with a
cubic superpotential ($k=2$ in \wpert); furthermore, we will only
discuss the special baryons with $n_1=N_f$, $n_2=N_c-N_f$ (see \baryon).
The main reason for considering these baryons is that they are dual by
\bmap\ to the simplest magnetic baryons, those with $\bar
n_1=2N_f-N_c=\bar N_c$ and $\bar n_2=0$.  Using \bmap\ and the phase $P$
found in the previous subsection we expect for these baryons the
following duality mapping:
\eqn\bartwo{
\eqalign{&B^{j_1\cdots j_{n_2}}\equiv
{1\over N_f!}\epsilon_{i_1\cdots i_{N_f}}
\epsilon^{\alpha_1\cdots\alpha_{N_f},\beta_1\cdots\beta_{N_c-N_f}}
Q^{i_1}_{\alpha_1}\cdots Q^{i_{N_f}}_{\alpha_{N_f}}
(XQ)^{j_1}_{\beta_1}\cdots
(XQ)^{j_{n_2}}_{\beta_{n_2}}=\cr
&{P\over\bar  N_c!}
{\bar s_0}^{N_f}\mu^{-\bar N_c}\Lambda^{2N_c-N_f}
\epsilon^{j_1\cdots j_{n_2},
\bar i_1\cdots\bar i_{N_f-n_2}}
\epsilon_{\bar\alpha_1\cdots\bar\alpha_{\bar N_c}}
q_{\bar i_1}^{\bar\alpha_1}\cdots q_{\bar i_{\bar
N_c}}^{\bar\alpha_{\bar N_c}}
\cr}
}
We will check the validity of \bartwo\ when a general superpotential
$W(X)={s_0\over3}\Tr X^3+{s_1\over2}\Tr X^2$ is introduced.
In this particular case $W$ has two critical points $a_1$, $a_2$,
and the electric and magnetic color groups are broken as in \dd, \ddd\
to: $SU(r_1)\times SU(r_2)\times U(1)$ and  $SU(\bar r_1)\times
SU(\bar r_2)\times U(1)$ respectively. The consistency check of \bartwo\
involves comparing two paths:

\item {1.} Decompose $B^{j_1\cdots j_{n_2}}$ in terms of the
electric baryons of SQCD and then use the SQCD mapping \barguru\ to
obtain an expression in terms of the magnetic SQCD baryons.

\item {2.} Use \bartwo\ first, and then decompose the magnetic baryon
into the SQCD baryons of the broken theory.

\noindent{}Clearly, the two expressions obtained this way should
coincide for agreement with duality. The electric baryon decomposes
after the breaking as:
\eqn\barel{\eqalign{
&B^{j_1\cdots j_{n_2}}={1\over
N_f!}\sum_{s=0}^{r_1}(-)^{(N_f-s)(r_1-s)} a_1^{r-s}
a_2^{N_f-N_c-(r_1-s)}
{N_f\choose s}{N_f-N_c\choose r_1-s}\epsilon_{i_1\cdots i_{N_f}}\cr
&\epsilon^{\alpha_1\cdots\alpha_{r_1}}Q^{i_1}_{1,\alpha_1}\cdots
Q^{i_s}_{1,\alpha_s}
Q^{j_1}_{1,\alpha_{s+1}}\cdots
Q^{j_{r_1-s}}_{1,\alpha_{r_1}}
\epsilon^{\beta_1\cdots\beta_{r_2}}Q^{i_{s+1}}_{2,\beta_1}\cdots
Q^{i_{N_f}}_{2,\beta_{N_f-s}}
Q^{j_{r_1-s+1}}_{2,\beta_{N_f-s+1}}\cdots
Q^{j_{n_2}}_{2,\beta_{r_2}}
}}
$Q_1$, $Q_2$ here are the SQCD quarks defined in section {\it 3.5}.
The non-trivial phases are due to the rearrangement of the $\epsilon$
symbol of $SU(N_c)$ in the order corresponding to a product of $\epsilon$
symbols of $SU(r_1)\times SU(r_2)$.
Next we replace the electric SQCD baryons in \barel\ by their magnetic
counterparts, using \barguru:
\eqn\barbie{
\eqalign{&\epsilon^{\alpha_1\cdots\alpha_{r_1}}Q^{i_1}_{1,\alpha_1}\cdots
Q^{i_s}_{1,\alpha_s}
Q^{j_1}_{1,\alpha_{s+1}}\cdots
Q^{j_{r_1-s}}_{1,\alpha_{r_1}}=\cr
&\epsilon^{i_1\cdots i_s j_1\cdots j_{r_1-s}
\bar i_1\cdots \bar i_{N_f-r_1}}\bar B^{(1)}_{\bar i_1\cdots \bar i_{
Nf-r_1}}
{1\over  ( N_f-r_1)!}\Lambda_1^{{1\over2}(3r_1-N_f)}\bar
\mu_1^{{1\over2}(N_f-r_1)}\cr}}
and an analogous relation for the baryon constructed out of $Q_2$.  The
scales $\Lambda_i$, and the parameters $\mu_i$ are given by
\lami, \blami. Another $s$ dependent phase (in addition to that in
\barel) appears {}from the rearrangement of the flavor indices in the
various $\epsilon$ tensors. After summing over the $i$ flavor indices
the sum over $s$ can be performed producing $(a_1-a_2)^{N_f-N_c}$.
Putting all the factors together we obtain:
\eqn\path{\eqalign{
&B^{j_1\cdots j_{n_2}}=(-)^{N_f-N_c}{N_f!\over
(N_f-r_1)!(N_f-r_2)!}\bar s_0^{N_f}\mu^{-\bar N_c}\Lambda^{2N_c-N_f}\cr
&\epsilon^{j_1\cdots j_{n_2},\bar j_1\cdots\bar j_{N_f-r_1},
\bar l_1\cdots \bar l_{N_f-r_2}}\bar B^{(1)}_{\bar j_1\cdots\bar
j_{N_f-r_1}} \bar B^{(2)}_{\bar l_1\cdots \bar l_{N_f-r_2}}\cr}}
It is straightforward to see that the same result for
$B^{j_1\cdots j_{n_2}}$ is obtained by decomposing the r.h.s.
of eq. \bartwo\ into baryons of $SU(\bar r_1)$ and $SU(\bar r_2)$.

\newsec{Examples.}

In this section we will study some of the consequences of the general
phenomena discussed in previous sections in some particular cases.

\subsec{The case $k=2$.}

 We start with the general cubic superpotential, \wpert\ with $k=2$.
The only coupling that exists in this case is $s_1$. For generic $s_1$,
\eqn\soo{W={s_0\over 3}\Tr X^3+{s_1\over 2}\Tr X^2}
$W$ has two critical points, $c_1$, $c_2$.  Vacua are labeled by
integers $r_1$, $r_2$ corresponding to placing $r_1$ of the eigenvalues
in the first critical point, and the remaining $r_2=N_c-r_1$ in the
second one.  For $r_1\not=0, N_c$ the gauge group breaks to (compare to
\dd):
\eqn\ktwob{SU(N_c)\to SU(r_1)\times SU(r_2)\times U(1).}
Eq. \wprime\ and the tracelessness
condition can be replaced by the two linear equations
\eqn\equ{\eqalign{c_1+c_2=&-{s_1\over s_0}\cr
                  r_1c_1+r_2c_2=&0\cr}}
with solution
\eqn\sol{\eqalign{c_1=&{s_1\over s_0}{r_2\over r_1-r_2}\cr
                  c_2=&{s_1\over s_0}{r_1\over r_2-r_1}\cr}}

The chiral ring is generated by $\Tr X^2$, whose expectation value in a
vacuum with multiplicities $(r_1,r_2)$ is:
\eqn\expec{\langle\Tr X^2\rangle=r_1c_1^2+r_2c_2^2=N_c\left({s_1\over2
s_0}\right)^2\left[{N_c^2\over(r_1-r_2)^2}-1\right]}
Similarly, in the magnetic theory:
\eqn\bsol{\eqalign
{\bar c_1=&{\bar s_1\over \bar s_0}{\bar r_2\over \bar r_1-\bar r_2}\cr
\bar c_2=&{\bar s_1\over \bar s_0}{\bar r_1\over \bar r_2-\bar r_1}\cr}}
and
\eqn\bexpec{\langle\Tr Y^2\rangle=\bar r_1\bar c_1^2+\bar r_2\bar c_2^2=
\bar N_c\left({\bar s_1\over2
\bar s_0}\right)^2\left[{\bar N_c^2\over(\bar r_1-\bar r_2)^2}-1\right]}
Comparing to \integ\ we see that to match the $r$ dependent terms in
$\langle\Tr X^2\rangle$ and $\langle\Tr Y^2\rangle$ we should choose
\eqn\maptwo{\bar s_1=s_1{N_c\over\bar N_c}}
which is a special case of \rels.
Comparing the constant terms of \expec, \bexpec\ then leads to
\eqn\mapmap{\alpha(s)={N_c\over24}{s_1^3\over s_0^2}\left({N_c^2\over\bar
N_c^2}-1\right)}
Eq. \mapmap\ is a special case of \nnn.
The operator relation \integ\ takes in this case the form:
\eqn\opop{\Tr X^2={N_c\over\bar N_c}\Tr Y^2+{N_c\over4}{s_1^2\over
s_0^2}\left({N_c^2\over\bar N_c^2}-1\right).}
which should be compared to \kk. There are two things to note here:

\item {1.} Considering the deformed theory with $s_1\not=0$ allows one
to uniquely determine the numerical coefficients in the operator mapping
\integ\ including the numerical constants $f_j$ \trx\ which are defined
in the theory with $s_1=0$.

\item {2.} While $c_i$ \sol\ and $\bar c_i$ \bsol\ depend on $r_i$,
they satisfy a simple linear relation
\eqn\roots{\bar c_i=c_i+{s_1 N_f\over s_0\bar N_c}.}
This linear relation is a special case of \oansatz.

For even $N_c$ we encounter here an example of an interesting general
phenomenon which occurs whenever $N_c$ and $k$ are not relatively
prime.  The solution for the eigenvalues $c_1$ and $c_2$ \sol\ is
singular when $r_1=r_2=N_c/2$ and $s_1 \not=0$ and therefore these vacua
are absent.  On the other hand, for $s_1=0$ the values of $c_1$ and
$c_2$ seem to be ambiguous.  Indeed, in this case the superpotential has
a flat direction with $\vev{X}={\rm diag}(c,...,c,-c,...,-c)$ (up to
gauge transformations).  The physics along this flat direction is
exactly that of the ``missing vacuum'' with $r_1=r_2=N_c/2$. The adjoint
field is massive and the gauge symmetry is broken:
$SU(N_c)\to\left[SU(N_c/2)\right]^2\times U(1)$.  The only difference is
that the mass of the adjoint field and the scales of the two SQCD
theories depend on the Higgs parameter $c$ instead of $s_1$.

How does the chiral ring look in this case? The only chiral operator in
the electric theory that can be written in terms of $X$ alone using the
equation of motion \h\ is $\Tr X^2$. The electric characteristic
polynomial gives rise to a relation for $X$ that depends on the parity
of $N_c$.  For odd $N_c$ this relation takes the form:
\eqn\oddnc{(\Tr X^2)^{N_c+1\over2}+b_1(\Tr X^2)^{N_c-1\over2}+\cdots+
b_{N_c-1\over2}(\Tr X^2)=0}
where $b_l$ are easily calculable coefficients which depend on
$s_1/s_0$. When $s_1=0$
all $b_l$ go to zero, and the relation is $(\Tr X^2)^{N_c+1\over2}=0$
in agreement with the fact that moduli space is in this case a point
(setting $M_j=0$ as usual). For generic $s_1$ \oddnc\ has $(N_c+1)/2$
solutions in one to one correspondence with the moduli space \expec.

For even $N_c$ the relation following {}from the characteristic
polynomial is
\eqn\evennc{s_1\left((\Tr X^2)^{N_c\over2}+ b_1(\Tr
X^2)^{N_c-2\over2}+\cdots+ b_{N_c\over2}(\Tr X^2)\right)=0.}
For non zero $s_1$ there are $N_c/2$ solutions in one to one
correspondence with the solutions of \expec. In particular, the solution
with $r_1=r_2$ which is singular does not appear. For $s_1=0$ the
characteristic polynomial \evennc\ does not provide any constraints on
$\Tr X^2$, in agreement with the presence of the flat direction
described above.

Quantum mechanically, we have learned that one needs to add to
\oddnc\ (or \evennc) another relation which is obtained
{}from the characteristic polynomial of the magnetic theory using the
operator map \opop. For odd $N_c$, $\bar N_c=2N_f-N_c$ is odd too, and
one finds a relation
\eqn\oddncbar{(\Tr Y^2)^{\bar N_c+1\over2}+b_1(\Tr Y^2)^{\bar
N_c-1\over2}+\cdots+ b_{\bar N_c-1\over2}(\Tr Y^2)=0}
There are two cases to discuss:

\item {1.} $N_c>N_f>\bar N_c$. In this case the electric theory is
more strongly coupled than the magnetic one. The relation \oddncbar\ is
non trivial in the electric chiral ring, reducing the number of distinct
vacua {}from $(N_c+1)/2$ to $(\bar N_c+1)/2$. This is in perfect
agreement with the counting of vacua \expec\ which satisfy $r_1, r_2\le
N_f$. The expectation values calculated {}from \oddncbar\ agree with
\expec\ for the appropriate vacua.

\item{2.} $\bar N_c>N_f> N_c$. The electric theory is more weakly coupled
than the magnetic one. The relation \oddncbar\ is satisfied on
all electric vacua satisfying the classical relation \oddnc\ and
therefore the electric moduli space is not modified quantum mechanically.
This is in agreement with weak coupling intuition in the electric
theory.

\subsec{The case $\bar N_c=kN_f-N_c=1$.}

For $k=1$ (SQCD with $N_c$ colors and $N_f=N_c+1$ flavors) the theory is
known to confine \natii. The low energy degrees of freedom are the
mesons $M^i_{\tilde j}$ and baryons $B_i$, $\tilde B^{\tilde j}$.  The
classical constraint relating these fields is lifted quantum
mechanically and is replaced by the superpotential:
\eqn\supqcd{W={1\over \Lambda^{2N_f-3}}\left(B_i M^i_{\tilde j}\tilde
B^{\tilde j}-\det M\right).}
This picture, which is due to strong coupling effects in the electric
theory, is much more simply understood in the weakly coupled dual
magnetic theory \nati. The mesons are the gauge singlets that appear in
the dual.  The baryons $B_i$, $\tilde B^{\tilde j}$ are proportional to
the magnetic quarks $q_i$, $\tilde q^j$.
The first term in the superpotential \supqcd\ is the tree level magnetic
superpotential (see e.g.\ \wmagsqcd). The last term (proportional to
$\det M$) is due to non-perturbative instanton effects that arise in the
process of complete breaking of the magnetic gauge group.

For $k>1$ we are in a position to use duality to make predictions.  The
operators generating the classical chiral ring of the $SU(kN_f-1)$ gauge
theory \b\ are $M_j$ \defM, $\Tr X^j$, $j=2,\cdots, k$, and the baryons
\baryon,
\eqn\bbbb{
\eqalign{
B^{(1)}_i=&Q^{N_f-1} (XQ)^{N_f}\cdots (X^{k-1}Q)^{N_f}\cr
B^{(2)}_i=&Q^{N_f} (XQ)^{N_f-1}\cdots (X^{k-1}Q)^{N_f}\cr
\cdots&\cdots\cr
B^{(k)}_i=&Q^{N_f} (XQ)^{N_f}\cdots (X^{k-1}Q)^{N_f-1}\cr}}
with $i=1,\cdots, N_f$, as well as antibaryons $\tilde B^{(j)}$.
However, unlike the case $k=1$,
a long distance description which includes all these fields does not
satisfy `t Hooft anomaly matching. Duality suggests another solution to
the problem.  The only independent fields in a macroscopic description
of the theory are $M_j$ and $B^{(k)}$, $\tilde B^{(k)}$.  The baryons
$B^{(k)}$, $\tilde B^{(k)}$ which we will denote by $B$, $\tilde B$ in
this subsection, are mapped by duality \bmap\ to the magnetic quarks.
The other generators of the chiral ring either vanish or can be
expressed in terms of $M_j$, $B$ and $\tilde B$. For example, {}from
\condd\ we deduce that:
\eqn\xxxx{{1\over k+1-i}\Tr X^{k+1-i}= \rho \tilde B M_{k-i} B+
{\partial\alpha_s\over\partial t_i}.}
where $\rho=-(-)^{{k(k-1)\over2}N_f}s_0^{-N_c-1}\Lambda^{1-(2k-1)N_c}$.
Clearly the chiral ring at long distances is drastically modified {}from
its classical structure. The superpotential of the theory can be read
off \wmagz:
\eqn\supp{W=s_0\rho\tilde B M_k B}
As a check note that the auxiliary scale $\mu$ present in \wmagz\
disappears in the electric variables $M$, $B$, $\tilde B$. The meson
fields $M_1$, $M_2$, $\cdots$, $M_{k-1}$ do not appear in the
superpotential. In general one expects additional dangerously irrelevant
terms in the superpotential \supp\ which do depend on all $M_j$ and
whose effect would be to lift some flat directions and
break some accidental symmetries. We have not analyzed these terms in
detail.

The theory has rather different descriptions at short distances, where
it is described in terms of gauge fields $W_\alpha$, and quarks $Q$,
$X$, and long distances where the quarks are confined and the
appropriate degrees of freedom are the mesons $M_j$ and baryons $B$,
$\tilde B$. It is natural to ask what is the long distance, macroscopic
description of deforming the superpotential $W$ by the $s_i$, as in
\wpert. The operator map \xxxx\ and deformed magnetic superpotential
\newwm\ with $Y$ set to $0$, tell us that in terms of the low energy
degrees of freedom the structure is always essentially the same. The
superpotential is
\eqn\superpo{W=\rho\tilde B\left(\sum_{l=0}^{k-1}t_lM_{k-l}\right)B}
The combination of mesons in brackets couples to the baryon field,
and the other $k-1$ combinations of generalized mesons remain free.

This is in agreement with the microscopic picture, according to which
turning on a generic superpotential \wpert\ leads to the appearance of
$k$ distinct critical points $a_i$ as described above. But because
$N_c=\sum_i r_i=kN_f-1$, there is a unique vacuum; we must assign
$N_f-1$ eigenvalues to one of the critical points (e.g. $r_1=N_f-1$) and
$N_f$ eigenvalues to each of the remaining $k-1$ ($r_i=N_f$ for
$i\ge2$). Thus, a microscopic physicist would expect to see at low
energies a set of almost decoupled SQCD theories, $k-1$ with $N_c=N_f$
and one with $N_c=N_f-1$, coupled by irrelevant interactions. According
to \natii, in the present context (remembering the gauged $U(1)^{k-1}$
-- see \dd), each of the theories with $N_f=N_c$ gives rise to a free
meson field, whereas the factor with $N_f=N_c+1$ gives rise to a meson
and baryon coupled via the superpotential \supqcd.

The combination of mesons in brackets in eq. \superpo\ corresponds to
the meson of the SQCD with $N_f=N_c+1$. The baryons $B$, $\tilde B$
correspond to the appropriate baryons of that theory. The other $k-1$
combinations of generalized mesons give the $k-1$ free mesons needed for
the theories with $N_f=N_c$. Comparing to \supqcd\ we learn that
\superpo\ is missing a term proportional to $\det\left(
\sum_lt_lM_{k-l}\right)$ which is allowed by symmetries and can therefore
appear much like in SQCD in the process of completely breaking the
magnetic gauge group. Its coefficient in the superpotential is
constrained by the symmetries to be a polynomial of the form
$t_{k-1}+\cdots + t_1^{k-1}$.  We leave the detailed
understanding of this term for future work.

It is not difficult to repeat the analysis above for the case when the
microscopic theory is deformed by \wpert\ with $s_i$ fine tuned such
that some of the critical points $a_i$ coincide. In fact, by requiring
the consistency of the resulting superpotential with all $\{s_i\}$
deformations puts extremely strong constraints on the corrections to the
superpotential \superpo\ and probably (over-) determines it. It would be
interesting to find this superpotential.

\subsec{The case $\bar N_c=kN_f-N_c=2$.}

The theory with $N_c=kN_f-2$ is the simplest theory with a non-Abelian
dual gauge group -- $SU(2)$.  Consider first the case $k=2$. The
electric theory has gauge group $G=SU(2N_f-2)$ and superpotential:
\eqn\wewe{W={s_0\over3}\Tr X^3.}
The magnetic one has gauge group $SU(2)$ and:
\eqn\wmwm{W_{\rm mag}={s_0\over\mu^2}\left(M_1\tilde q Y q+
M_2\tilde q q\right).}
The magnetic theory has no superpotential for $Y$ (setting other
fields to their expectation values). Thus, there is a flat direction
corresponding to
\eqn\mfl{\langle Y\rangle=a
\left(
\matrix{1&0\cr
0&-1\cr}
\right).}
Along it the gauge group is broken: $SU(2)\to U(1)$. $Y$ is massive
except for one massless field, $\Tr Y^2$, parametrizing small
fluctuations of the expectation value $a$.

Remarkably, the electric theory also has a flat direction, since $N_c$
(which is even) is divisible by $k$ ($=2$).  This flat direction, which
was discussed in {\it 6.1} is described by:
\eqn\efl{\langle X\rangle={\rm diag}\left(b^{N_f-1},
(-b)^{N_f-1}\right).}
It too breaks the (electric) gauge symmetry:
\eqn\brga{SU(2N_f-2)\to SU(N_f-1)\times SU(N_f-1)\times U(1).}
$X$ is massive except for one massless field, $\Tr X^2$,
parametrizing small fluctuations of $b$. The non-Abelian dynamics in
\brga\ is SQCD with $N_f=N_c+1$. We expect two sets of mesons and
baryons coupled by the superpotential \supqcd. Substituting \mfl\ in
\wmwm\ we indeed find the required structure.  Linear combinations of
the two meson fields $M_1$, $M_2$ couple to the two components of $q,
\tilde q$ which serve as baryons.  The couplings look like the first
term in \supqcd. We are again not analyzing the $\det M$ terms due to
instantons that must be added to the magnetic superpotential \wmwm\ in
the weakly coupled magnetic theory.  The $U(1)$ dynamics also agrees and
the massless field $\Tr X^2$ corresponds to $\Tr Y^2$ completing the
duality map.

At the origin ($a=b=0$ in \mfl, \efl) we find a strongly coupled
dual of the $SU(2)$ theory with $W=0$.

For even $k>2$ the structure is similar but richer.  Theory space has
many components depending on the number of distinct critical points.
Consider, for example, the case when all $k$ critical points are
distinct. Since we want the magnetic superpotential for $Y$ to vanish we
add only odd powers of $X$ to the electric superpotential \wpert\ (i.e.\
all $s_{2l+1}$ vanish). Therefore, the critical points appear in pairs,
$\pm a_1$, $\pm a_2$, $\cdots$, $\pm a_{k\over2}$. There is a unique
vacuum with
\eqn\brbr{SU(kN_f-2)\to \left[SU(N_f-1)\right]^2\times
\left[SU(N_f)\right]^{k-2}.}
The magnetic $SU(2)$ gauge theory is coupled to $k$ singlet fields $M_j$,
$j=1,\cdots, k$ and
$$W_{\rm mag}=\sum_{j=1}^k M_j\tilde q Y^{k-j} q.$$
There is still a flat direction \mfl, along which $k-2$ combinations
of $M_j$ do not couple (in the IR at the origin of moduli space
$\langle M_j\rangle=0$), while two combine as before with $q, \tilde q
\to B, \tilde B$ to give the right structure for two SQCD theories with
$N_f =N_c+1$. The non-renormalizable part of the superpotential again
remains to be analyzed. It is not hard to repeat these considerations
for the case when some eigenvalues $a_i$ coincide. We leave the details
to the reader.

It is appropriate to end this section by returning to the truncation
of the chiral ring, and explain why the quantum chiral ring is smaller
than the classical one. In the example discussed here, in the quantum
chiral ring there are relations:
\eqn\relat{\Tr X^{2n+1}={\rm const};\;\;\; \Tr X^{2n}=F(\Tr X^2)}
which are absent classically.

Consider first the classical theory where as usual we deform $W$ \wpert\
to resolve the singularity. One way of seeing that $\Tr X^j$,
$j=2,\cdots, k$ are all independent classically is to note that the
theory with generic $\{s_i\}$ \wpert\ has many classical vacua in which
$\langle \Tr X^j \rangle$ are all different. There are no relations
among the $\langle\Tr X^j\rangle$ that hold uniformly in all vacua. The
key point is that as discussed previously many of the classical vacua
are unstable and disappear in the quantum theory, due to the $N_f\ge
N_c$ constraint in the low energy SQCD theories (see section {\it 2}).
As we saw, in the case $N_c=kN_f-2$ analyzed here most of the vacua
disappear, leaving behind a unique vacuum (for generic
$s_i$).  Relations among the operators need to hold only in the quantum
vacua, and therefore many more exist in general quantum mechanically
than classically.  One can easily convince oneself that the relations
\relat\ in particular are valid in all the vacua that are stable quantum
mechanically and break down in vacua that are unstable. Therefore they
only exist in the quantum theory and not in the classical one.

\newsec{Comments on the theory without a superpotential}

The $SU(N_c)$ theory with an adjoint $X$, $N_f$ fundamentals $Q$ and
$N_f$ anti-fundamentals $\tilde Q$ without a tree level superpotential
is very interesting.  Unfortunately, our understanding of this theory is
very limited.  In the previous sections we followed \refs{\dk, \ks} and
added the tree level superpotential ${1 \over k+1} s_0 X^{k+1}$ to
simplify the analysis.  In this section we will point out a few
observations about the theory without the superpotential ($s_0=0$).

\subsec{The moduli space}

No dynamically generated superpotential can lift the classical flat
directions.  To see that, note that at a generic point in the moduli
space the gauge group is broken to an Abelian subgroup (or completely
broken).  Therefore, instanton methods should be reliable.  Repeating the
analysis of \ads, it is easy to see that instantons have too many fermion
zero modes to generate a superpotential.  Therefore, no superpotential
can be generated and the theory has a moduli space of stable vacua for
any $N_f$.  The same conclusion can be reached by examining the
symmetries and the way they restrict a dynamically generated
superpotential.

Furthermore, the perturbations of this theory by various tree level
superpotentials (see below) relate it to other theories where the moduli
space is understood.  This leads to further constraints on the quantum
moduli space and its singularities.  The conclusion is that the quantum
moduli space is identical to the classical one.  The only singularities
are at points where classically the unbroken gauge symmetry is enhanced.
The most singular point is at the origin.  This does not mean that the
physical interpretation of the singularities in the quantum theory is as
in the classical theory.

\subsec{The non-Abelian Coulomb phase $N_f \roughly< 2N_c$}

When $N_f \roughly< 2N_c$ a simple two loop calculation similar to that
in \bankszaks\ reveals a non-trivial fixed point of the
beta function.  This calculation can be justified rigorously at large
$N_c$.  Therefore, the origin is in a non-Abelian Coulomb phase.  The
physics at that point is similar to the physics seen in perturbation
theory, describing interacting quarks and gluons.

The dimensions of the chiral operators are constrained by the
superconformal algebra to satisfy $D={3 \over 2 }R$ where $R$ is the
charge of the $U(1)_R$ symmetry in the superconformal algebra.  In
simpler examples this $U(1)_R$ symmetry is easily identified \nati.  In
our case, it is ambiguous.  Anomaly freedom constrains the $R$ charge of
$Q$ and $\tilde Q$, $B_f$ and the $R$ charge of $X$, $B_a$ to satisfy
\an\ $N_f B_f+N_c B_a=N_f$.  For $\epsilon={2N_c-N_f \over 2N_c} \ll 1 $
the fixed point is visible in perturbation theory with the conclusion
$B_a={2\over 3}(1- \epsilon {N_c^2\over 2 N_c^2-1} + \CO(\epsilon^2))$;
$B_f={2\over 3}(1- {\epsilon \over 2}{N_c^2-1 \over 2N_c^2-1} +
\CO(\epsilon^2))$.

As $N_f$ is reduced, the fixed point becomes more strongly coupled and
the dimensions of the operators become smaller.  Since the dimensions of
the the spin zero fields cannot be smaller than one
\ref\mack{G. Mack, \cmp{55}{1977}{1}},
at some point this description must be modified.  We would expect, by
analogy with \nati, that these fields become free and decouple.  Then,
the $U(1)_R$ in the superconformal algebra is an accidental symmetry
which assigns $R={2 \over 3}$ to these fields.
This observation might suggest that all $u_j=\Tr
X^j$ with $j=2,...,N_c$ should be included
as elementary fields in a dual description.

\subsec{The confining phase superpotential}

A useful tool in analyzing the dynamics of supersymmetric theories is
the confining phase superpotential.  It is obtained by coupling the
generators of the chiral ring to external sources and computing the
effective action for these sources.  Upon a Legendre transform this
leads to an effective action for these composites which gives a good
description of the confining phase of the theory
\nref\isphases{K. Intriligator and N. Seiberg, \np{431}{1994}{551}.}%
\refs{\isphases,\lectures}.  This procedure was used in \isphases\ to
study the $N_c=2$, $N_f=1$ theory.  The authors of
\ref\jerusalem{S. Elitzur, A. Forge, A. Giveon
and E. Rabinovici, \pl{353} {1995} 79, hep-th/9504080; hep-th/9509130.}
studied the $N_c=2$ problem for larger values of $N_f$ and gave partial
answers for larger $N_c$.  The massless modes at the generic point in
the moduli space are among these fields and therefore this effective
superpotential gives a good description of the theory on the moduli
space.  The effective superpotential derived this way exhibits a
singularity at the origin for every value of $N_c$ and $N_f$.  This
means that at the origin more degrees of freedom are needed.  This fact
is in accord with our interpretation of the origin as being in a
non-Abelian Coulomb phase.  However, we could not find a useful
description of the theory at the origin which gives rise to the
singularity in these effective superpotentials.

\subsec{The Coulomb phase}

The theory with a tree level superpotential $\lambda_i^{\tilde i}
\tilde Q_{\tilde i} X Q^i= \Tr \lambda M_2$ can be analyzed easily.  For
$\lambda_i^{\tilde i}= \delta_i^{\tilde i}$ the theory becomes $N=2$
supersymmetric.  This theory for $N_c=2$ was analyzed in \swii\ and for
larger values of $N_c$ in
\nref\ntwomatti{A. Hanany and Y. Oz,
hep-th/9505075, \np{452} {1995} 283.}%
\nref\ntwomattii{P.C. Argyres, M.R. Plesser and A. Shapere,
hep-th/9505100, \prl{75} {1995} 1699.}%
\nref\aps{P.C. Argyres, M.R. Plesser and N. Seiberg, to appear.}%
\refs{\ntwomatti - \aps}.  The moduli space of the theory has a Coulomb
branch which has only massless photons at generic points.  At special
singular points on the moduli space there are more massless particles:
massless monopoles, massless dyons, massless gluons and quarks, and even
points with interacting non-trivial $N=2$ superconformal field theories
\nref\AD{P. Argyres and M. Douglas, hep-th/9505062, \np{448} {1995} 93.}%
\nref\apsw{P.C. Argyres, M.R. Plesser, N. Seiberg and E. Witten, to
appear.}%
\refs{\AD, \apsw}.  More quantitatively, this branch of the theory is
described in terms of a hyperelliptic curve.  The characteristic scale
on this Coulomb branch is the only dimensionful parameter in the theory
$\Lambda$ which appears as a parameter in the curve.

It is easy to extend the curve away from this $N=2$ theory; i.e.\ for
arbitrary values of $\lambda_i^{\tilde i} \not= \delta_i^{\tilde i}$.
As in \refs{\isphases, \jerusalem,\ntwomatti} using the symmetries of
the theory, this is achieved by replacing every factor of
$\Lambda^{2N_c-N_f} $ in the curve by $(\det \lambda)
\Lambda^{2N_c-N_f}$.  Therefore, as $\lambda \rightarrow 0$, all
the features on the Coulomb branch approach the origin (more precisely,
they approach points where classically there is an unbroken non-Abelian
gauge symmetry).   Conversely, by turning on $\lambda$ the singularity
at the origin splits to several singularities with various massless
particles.  Since these particles are not all local with respect to one
another, there is no local Lagrangian which includes all of them.
Therefore, it is impossible to write a local field theory which
describes the deformation by $\lambda$ along the entire Coulomb branch
in weak coupling.

A similar situation was encountered in $SO(N_c)$ gauge theories
\nref\isprev{K. Intriligator and N. Seiberg, hep-th/9506084, to appear
in the Proc. of Strings '95.}%
\refs{\isson,\isprev}.  There the theory at the origin was given several
different dual descriptions.  Each gave a weak coupling description of
another deformation or another region of the moduli space.  An attempt
to imitate this procedure here will necessarily lead to a very large
number of dual theories to describe the different phenomena in the
Coulomb branch.

\subsec{Deformation by a superpotential ${s_0 \over k+1}\Tr
X^{k+1}$}

This is the theory we studied in this paper.  Removing this perturbation
by letting $s_0$ go to zero is a singular operation as it changes the
asymptotic behavior of the potential.  This can also be seen {}from the
matching equation $\Lambda^{2N_c-N_f} \bar \Lambda^{2\bar N_c-N_f} =
\left({\mu\over s_0}\right)^{2N_f} $ which becomes singular as $s_0$
goes to zero.  Alternatively, we might attempt to remove this
perturbation by letting $k$ go to infinity.

As $k$ becomes large, the operator $\Tr X^{k+1}$ becomes irrelevant at
the long distance theory at the origin and therefore it does not affect
the dynamics.  As we noted above, this operator is dangerously
irrelevant and cannot be ignored.  However, as $k$ goes to infinity the
potential it leads to becomes very flat and it is it reasonable that the
theory without a superpotential is achieved.

For large $k$ the dual gauge group $SU(\bar N_c= kN_f-N_c)$ becomes
large.  This theory is strongly coupled and therefore one might think
that it does not lead to a useful dual description.  However, as we saw
in the previous sections, this theory gives a weak coupling description
of some of the deformations.  Therefore, if there is a unique good dual
theory at the origin, it should include this $SU(\bar N_c)$ for
arbitrarily large $\bar N_c$.  Such an $SU( \infty)$ theory is expected
to behave like a string theory.  Therefore, we might speculate that the
dual theory at the origin is not a field theory but a string theory.
Perhaps, if this is indeed the case, there will not be a need for a
large number of dual descriptions as suggested by the structure of the
Coulomb phase.

\bigskip

\centerline{{\bf Acknowledgments}}

We would like to thank T. Banks, K. Intriligator, S.  Shenker, and
especially E. Witten for many helpful discussions.  This work was
supported in part by DOE grant \#DE-FG05-90ER40559, BSF
grant number 5360/2, the Minerva foundation and a DOE OJI grant.
A.S. would like to thank the hospitality of the Enrico Fermi Institute
where part of this work was performed. D.K. thanks the Weizmann Institute,
Aspen Physics Center and Department of Physics at Rutgers University for
hospitality during the course of this work.

\appendix{A}{Dangerously irrelevant operators}

A quantum field theorist might wonder about the presence of high order
polynomials such as \b\ in the superpotential.  These are
non-renormalizable interactions which seem irrelevant for the long
distance behavior of the theory.  How is it then that they affect the
physical results?  In order to answer this question we should review the
notion of a dangerously irrelevant (DI) operator in the theory of the
renormalization group.

Consider deformations of a fixed point of the renormalization group by
relevant operators or along flat directions of the potential.  It might
be that an irrelevant operator at the original fixed point becomes
relevant after the deformation.  Such an operator is called dangerously
irrelevant; ignoring its presence will lead to incorrect results.

An example of a dangerously irrelevant operator is the gauge superfield
$W_\alpha W^\alpha$ in a non asymptotically free gauge theory (for
definiteness one may think of supersymmetric QCD with quarks in the
fundamental representation of the gauge group). In such a situation the
gauge coupling flows to zero at long distance; hence the operator
$W_\alpha W^\alpha$ is irrelevant. Nevertheless, it is clearly important
to keep the gauge coupling when describing the long distance behavior of
gauge theories. If, for example, we turn on quark masses, the number of
light quarks may fall below the asymptotic freedom bound and therefore
the gauge coupling becomes relevant.  Thus, $W_\alpha^2$ is relevant in
part of theory space and irrelevant in other parts. When it is
irrelevant it is referred to as being dangerously irrelevant.

The operator $\Tr X^{k+1}$ behaves in a very similar way.  At the
gaussian (UV) fixed point it has (for $k>2$) dimension larger than
three.  However, if we turn on the gauge coupling $g$, $\Tr X^{k+1}$
develops for sufficiently small $N_f$ an anomalous dimension which can
make it relevant. Hence, while this operator is irrelevant near the
gaussian UV fixed point, it too is a dangerously irrelevant operator.

Actually, $\Tr X^{k+1}$ is dangerously irrelevant even without the gauge
coupling. While it is irrelevant when expanding around the trivial,
$X=0$ vacuum, it is clearly relevant when expanding around any non-zero
$X$. Thus, the presence of the superpotential \b\ lifts some of the flat
directions of the original theory; also, turning on a polynomial
superpotential \welec\ one finds many minima at non vanishing $X$, where
clearly all powers up to and including $\Tr X^{k+1}$ are relevant.
Therefore, at the origin $\Tr X^{k+1}$ is a DI operator and cannot be
ignored.

{\appendix{B}{Properties of symmetric polynomials.}

Consider a polynomial of order $k+1$,
\eqn\aaa{W(x)=\sum_{i=0}^k{1\over k+1-i} s_ix^{k+1-i}}
The $k$ roots of $W^\prime(x)$, $a_i$ ($i=1,\cdots k$) satisfy:
\eqn\bbb{W^\prime=\sum_{i=0}^k s_i x^{k-i}\equiv
s_0\prod_{i=1}^k(x-a_i)}
where $\{s_l\}$ and $\{a_i\}$ are related by:
\eqn\ccc{s_l=(-)^ls_0\sum_{i_1<i_2<\cdots<i_l} a_{i_1}a_{i_2}\cdots
a_{i_l}}
It is also natural to define the objects:
\eqn\d{u_l\equiv \sum_{i=1}^k a_i^l}
which satisfy the recursion relation:
\eqn\eee{ls_l+\sum_{i=1}^ls_{l-i}u_i=0;\;\;\;l=1,2,3,\cdots}
Eq. \eee\ can be thought of as determining $u_l$ in terms of $\{s_m\}$
with $m\le l$. For the few lowest cases we have:
\eqn\lowu{
\eqalign{
u_1=&-{s_1\over s_0}\cr u_2=&\left({s_1\over
s_0}\right)^2-2\left(s_2\over s_0\right)\cr u_3=&3{s_2\over
s_0}{s_1\over s_0}-\left(s_1\over s_0\right)^3-3{s_3\over s_0}\cr }}
etc. In the text (section {\it 3}) we used a remarkable property of the
$u$'s:
\eqn\newt{
{\partial\over\partial s_j}{u_{j+l}\over j+l}=
{\partial\over\partial s_i}{u_{i+l}\over i+l}}
For all $j, l$.
In other words, duality requires that ${\partial\over\partial
s_j}{u_{j+l}\over j+l}$ should be independent of $j$.
The proof of this statement relies on the following representation of the
recursion relation \eee:
\eqn\ident{\sum_n {t^n\over n}u_n=\ln\left(1+\sum_is_it^i\right)}
Differentiating \ident\ w.r.t. some $s_j$ we find:
\eqn\identit{\sum_n {t^{n-j}\over n}{\partial\over\partial s_j}u_n=
{1\over 1+\sum_is_it^i}}
which makes the fact that ${\partial\over\partial s_j}u_n/n$ depends
only on $n-j$ and not on $n$, $j$ separately manifest.}

\listrefs
\bye